\documentclass[journal,12pt]{IEEEtran}

\usepackage{ifpdf}
\usepackage{cite}
\usepackage{graphicx}
\graphicspath{{./}}
\DeclareGraphicsExtensions{.pdf,.jpeg,.png}
\ifCLASSINFOpdf
\else
\fi
\usepackage[cmex10]{amsmath}
\usepackage{amssymb}
\usepackage{wasysym}
\usepackage{algorithm}
\usepackage{algpseudocode}% http://ctan.org/pkg/algorithmicx
\usepackage{array}
\usepackage{breqn}
\usepackage[inline]{enumitem}

  \usepackage[caption=false,font=footnotesize]{subfig}
\usepackage{stfloats}
\usepackage{url}
\usepackage{setspace}
\usepackage[dvipsnames]{xcolor}

\newcommand{\R}{\mathbb{R}}

\newcommand*\mean[1]{\bar{#1}}
\begin{document}

\title{\singlespacing fCDN: A Flexible and Efficient CDN Infrastructure without DNS Redirection or Content Reflection}

\author{\singlespacing Mays~Al-Naday,~\IEEEmembership{Member,~IEEE,}
        Martin~J.~Reed,~\IEEEmembership{Member,~IEEE,}
        Janne~Riihij\"arvi,~\IEEEmembership{Member,~IEEE,}
        Dirk Trossen,~\IEEEmembership{Member,~IEEE}
        Nikolaos Thomos,~\IEEEmembership{Senior Member,~IEEE,}
        and~Mohammed~Al-Khalidi,~\IEEEmembership{Student Member,~IEEE}
        % <-this % stops a space
%\thanks{Mays~Al-Naday, Martin Reed, Nikos Thomos and Mohammed~Al-Khalidi are with the School of Computer Science and Electronic Engineering, University of Essex, Colchester, UK, CO4 3SQ,
%(e-mail: \{mhfaln,mjreed,nthomos,mshawk\}@essex.ac.uk).}% <-this % stops a space
%\thanks{Janne~Riihij\"arvi is with Institute for Networked Systems, RWTH Aachen University, 52072 Aachen, Germany
%(e-mail: jar@inets.rwth-aachen.de)}% <-this % stops a space
%\thanks{Dirk Trossen is with InterDigital Europe, Ltd. London, EC2A 3QR, UK (e-mail: dirk.trossen@interdigital.com)
%}%
}

\maketitle
\vspace{-2\baselineskip}
\begin{abstract}
Flexible and efficient CDNs are critical to facilitate content distribution in 5G+ architectures. Current CDNs suffer from inefficient request mapping based on DNS redirection, and inefficient content distribution from origin to edge servers, through content reflection.
We proposes a novel, flexible CDN architecture that removes the need for DNS-based mapping and content reflection. 
Instead, requests to/from the CDN are treated as service transactions in the network, which utilises a routing function embraced from emerging research in Information-Centric Networks (ICN) to route edge-to-edge transactions to the true nearest service point. The same function is utilized to establish path-based flows over a fast forwarding substrate; thereby, eliminating the need for IP routing between service points within a single domain, and potentially at peering points with other domains. 
%Our solution decouples services from location and allows for fine-grained and fine-tuned service transactions, while increasing the efficiency in utilising the network resources.
 We model our architecture and formulate the resource placement problem as a variance of the ``$K$-center'' problem. To address the problem, we propose a greedy algorithm, \emph{Swing}, that balances the placement of service points between highly and poorly connected nodes. We evaluate the efficiency of our architecture in utilising the CDN and network resources through Monte Carlo simulations that explore a range of $K$ values. Moreover, we compare the goodness of the placement solutions provided by Swing with those provided by \emph{Largest First} and \emph{Closest First} Algorithms. Evaluation results show the superiority of our fCDN solution in reducing the edge-to-edge path length and the required network resources.

\end{abstract}

% \begin{IEEEkeywords}
% CDN, DNS, ICN, Mapping Overlay, Transport Overlay.
% \end{IEEEkeywords}

% \fi
\IEEEpeerreviewmaketitle

\section{Introduction}
\IEEEPARstart{C}{ontent} Distribution Networks (CDN) emerged as application overlays that provide information awareness on top of the existing IP architecture. They facilitate delivering live and on-demand multimedia content to customers, by localising content nearer to the end-user. Current CDNs require solutions such as DNS (and HTTP) redirection and content \emph{reflection} to enable large scale content distribution. This is because existing network architectures strongly tie information to location, namely in the Domain Naming Service (DNS) and the routing/forwarding functions. However, DNS redirection and content reflection limit the flexibility of CDNs and increase the overlay complexity. This means that existing CDN solutions do not easily accommodate 5G+ edge architectures, which aim to reduce latency for content retrieval and improve routing efficiency to accommodate the anticipated growth in network capacity requirements. The need to address these issues is eminent in 5G+ networks, in order to support the expected shift in communications towards mobile services, where CDN is going to be pushed increasingly towards the edge.

Traditional CDNs are implemented through three intertwined overlays: the (user) mapping overlay, which directs the user request to the most appropriate CDN Point of Presence (PoP); the transport overlay, which disseminates content from the origin server(s) to the edge servers, such as for live-streaming; and, the routing overlay, used for the first two overlays. These overlays~\cite{Sitaraman2014,Chen2012,Papagianni2013} are particularly visible in an Akamai~\cite{Sitaraman2014} CDN, with the mapping overlay being responsible for assigning CDN PoPs to Local DNS (LDNS) points; and, the transport overlay facilitating edge-origin communication through application-layer multicast. This paper presents a novel CDN architecture that provides the features of these overlays while overcoming the limitations of CDNs by introducing a service routing architecture that allows greater edge integration.

Today, user mapping through request redirection is the main solution used by CDNs to provide \emph{anycast}. DNS redirect relies on maintaining multiple IP addresses to a single fully qualified domain name (FQDN) and ``mapping'' the FQDN to one of the IP addresses, depending on the location of the IP addresses. DNS redirect is the cheapest form of the redirection, as it only depends on the DNS server location and incurs much less overhead than other forms of redirect, such as the HTTP counterpart. However, this mechanism relies on selecting a ``nearest'' PoP to the LDNS point, not to the client.
%based on the client's choice of DNS server without any accurate topological information. 
This results in some scenarios having highly inefficient mappings; thereby, degrading the end-to-end QoS while increasing load inside the network. Additionally, DNS was not designed to operate in a highly flexible manner to deliver agile service routing. 
%These issues have been addressed with an end-user extension of the DNS protocol \cite{Chen2015}, whereby an IP prefix is used to determine the subnet from which the request originated and thus select the nearest PoP. However, the solution provides a one-to-one mapping, with the solution quality degrading as the network diameter increases.

Once a user request is mapped, the transport overlay comes to bear and is particularly evident in streaming CDNs. Today, the transport overlay introduces a middle substrate of content ``splitters'', namely \emph{reflectors} which scale up the dissemination of content from \emph{entry points} (or originators) to edge servers. This is enabled through a form of application-layer multicast where a reflector acts as the splitting and replication point. Notably, the reflector substrate mainly compensates for the lack of scalable anycast/multicast support at the network layer. Therefore, it can be thought of as an expensive addition to provide a feature that ought to be natively supported by the network. This not only increases the network overall cost, by the added cost of the middle substrate of reflectors, but also results in sub-optimal routing between edge servers and origins through the use of unicast routing for traffic that could otherwise be transmitted using multicast, and through the inability of existing IP routing to adequately separate location from the service name.

This paper proposes a novel, flexible, CDN architecture for matching supply with demand, that eliminates DNS redirection and content reflection from CDN, while allowing for flexible, and highly agile, composition of anycast/multicast relationships; therefore, aiming to positively impact user-level latency as well as network utilization. 
This is achieved by interpreting user requests to CDNs, as well as the transport from the origin server to CDNs, as a service transaction. We employ a novel service routing solution, directly realized over a L2 transport network. Such service routing eliminates the need for DNS and HTTP redirection, whilst utilizing an efficient, agile, L2 multicast solution for transport in the network and therefore removing the need for reflectors. In our evaluation, we focus on the efficiency gains of our proposed mapping%, achieved by using our ICN routing function
%removing the need for DNS-based indirection
, leaving the improvements of the transport for future work. 
 For our realization, we utilize insights from the emerging research in Information-Centric Networking (ICN) that natively supports \emph{anycast}, while path calculation and forwarding is facilitated through source-routing mechanisms that natively support multicast in the network.
 Whilst our realization is rooted in first deployments of real prototypes in early trials, our ideas are supported by first contributions to standardization, most prominently in the IETF. For this, we relate our ideas to that of Service Function Chaining (SFC), albeit elevating the service routing to the level of HTTP, leaving the routing decisions to a substrate similar to the one outlined in this paper. Within the SFC WG of the IETF, contributions to this effect are currently being discussed in~\cite{ietf2017}. An early realization of the service routing substrate has also been part of an ETSI Multi-access Edge Computing (MEC) proof-of-concept~\cite{mec2017},
%  [https://mecwiki.etsi.org/index.php?title=PoC\_4\_FLIPS\_\%E2\%80\%93\_Flexible\_IP-based\_Services]
which has showcased efficient content delivery for HTTP-level streaming as well as edge content retrieval in localized gaming.

Our contributions in this paper are threefold:
\begin{enumerate*}
\item First, we describe the design of a novel, flexible CDN architecture, (\emph{fCDN}) freed from DNS redirection and content reflection; and, instead, comprising a set of functions focused on providing content naming and CDN resource management. We present the CDN consumption and service points as edge clusters that communicate over a fast P2MP core. We describe the network function utilized in the core to provide edge-to-edge service routing and show how it can efficiently replace DNS redirection in providing \emph{anycast}.
\item Second, we define the model to describe our proposed architecture and formulate the resource placement problem as a variance of the ``$K$-center'' problem.
%\item Second, we define the mathematical model to describe the relationships between publications and subscriptions, and the resultant traffic in the fCDN. We use this model to define the resource requirements in the network, in terms of storage and link capacities; and formulate the resource placement problem as a variance of the ``$K$-center'' problem. 
\item Third, propose a novel greedy algorithm, namely \emph{Swing} a variance of the Farthest First Travel algorithm to reduce edge-to-edge path; thereby, allowing for reducing the network requirements and end-user latency. We develop Monte Carlo simulations that explore values of $K$ in Swing, and use that to identify a suitable parameter set that minimises the overall storage and network requirements.
\end{enumerate*}

The rest of this paper is structured as follows: Section~\ref{sec:related_work} outlines the inefficiencies in mapping requests based on DNS redirection, and the user-extension to DNS adopted to address these inefficiencies. Section~\ref{sec:cdn_arch} presents our proposal for a flexible CDN and describes how the network functions, the service routing required at the edge and the CDN functions. Section~\ref{sec:model_and_problem} defines the mathematical model for placing and delivering CDN services, and Formulates the optimization problem for resource placement. Section~\ref{sec:algo} presents our greedy algorithm, Swing, and the reference algorithms compared against it. Section~\ref{sec:eva} presents the performance evaluation of our proposed fCDN and the greedy algorithm Swing, and finally Section~\ref{sec:conclusion} draw our conclusions.
%The mechanism requires the CDN naming authority - part of the mapping overlay - to associate each Local DNS point (LDNS), in an operators network, with the nearest one or more CDN Points of Presence (PoPs). Once associated, those PoPs becomes the content \emph{providers} to all the clients contacted by the respective LDNS. 

\section{Related Work}
\label{sec:related_work}
% In an operator’s network, a CDN may either be deployed by the operator, namely network-based CDN; or it can be offered by a 3rd party CDN provider, known as independent CDN.
% Today, a large number of CDNs co-exist and peer with each other in an operator's network and cross networks; most of which comprized from proprietary solutions and strictly non-disclosed design choices to enable the scalability objectives of the CDN provider. Nevertheless, 
Existing research shows that, in general, major CDN architectures, such as Akamai~\cite{Sitaraman2014,Kontothanassis2004} and Google~\cite{Krishnan2009,Rafetseder2011}, are comprized of a set of overlays, particularly: the \emph{mapping}, \emph{transport} and \emph{routing} overlays. 
We focus here on the mapping overlay, which relies generally on DNS redirection of requests to the nearest CDN PoP. To realise this behaviour in an operator's network, the mapping overlay assigns CDN PoPs to LDNSes; whereby, the nearest PoP(s) is assigned to each LDNS point. Notice here that \emph{nearest} refers to the distance between the LDNS and the CDN PoP, not to the distance between end-users access nodes and the PoP. This means that in a scenario where a CDN PoP is located in an access node, but the LDNS point is not co-located in the same nodes, end-user requests will not be mapped to this nearest CDN PoP. Instead, end-user requests will be mapped inefficiently to remote PoPs. The severity of this issue has been quantified by~\cite{Chen2015} for public DNS resolvers; and, an extension to the DNS protocol to include the IP prefix of the client network has been proposed and deployed in major CDNs, such as that of Akamai and Google~\cite{Chen2015}. The solution of~\cite{Chen2015} only mitigates the problem, as it only provides one to one mapping between a subnet and a CDN PoP. Moreover, given the sharp increase in the size of the DNS state, the length of the IP prefix has to constrained, such that the DNS state remains manageable. A general agreement on $/20$ has been conveyed in~\cite{Chen2015}, which translates into a subnet of up to $4094$ users. While this has been a reasonably small subnet in existing networks; the push for mobile edge communications in 5G+ architectures is expected to demand considerably higher granularity.

The massive progress of CDN has paved the away for emerging research in Information-Centric Networking (ICN). ICN advocates for rethinking the design of the Internet to overcome the limitations mainly faced by CDN overlays~\cite{Fotiou2010,Zhang2014}. However, the ICN proposition to radically change the semantics of the communication paradigm, on a global scale, makes it considerably challenging to adopt ICN by operators; because, this requires massive update to current and legacy infrastructure. Migration approaches towards deployable ICN has been proposed in~\cite{Fayazbakhsh2013,Trossen2015}. The proposition of~\cite{Fayazbakhsh2013} is to deploy ICN on top of existing CDNs; thereby, inheriting the inefficiencies of DNS-based redirection and the suboptimal transport through reflectors, discussed in the Introduction. In contrast,~\cite{Trossen2015} proposes to maintain IP services at the edge of an operator's network; while facilitating edge-to-edge communications a L2-like fabric in the core. The latter proposition paves the away for overcoming the inefficiencies of DNS redirection and content reflection, as we will show when describing our proposal in Section~\ref{sec:cdn_arch}. Next, we overview the proposition of~\cite{Trossen2015} and describe the ideas we use from this for our proposal for flexible CDN.

% Nevertheless, existing research shows general design guidelines that are followed by major CDN providers, such as Akamai~\cite{} and Google~\cite{} 
% with   In an operator’s network, a CDN may either be deployed by the operator, namely network-based CDN; or it can be offered by a 3rd party CDN provider, known as independent CDN. 
\subsection{IP-over-ICN}
\label{sec:ip_over_icn}
IP-over-ICN architecture has emerged as a feasible migration path towards ICN~\cite{Trossen2015,Reed2016,ALNaday2017}. This is now moving from a research idea towards standardization as described in the Introduction~\cite{ietf2017,mec2017}. An IP-over-ICN architecture assumes an operator's network to consist of two parts: a set of IP networks at the edge; and, a Point-to-Multi-Point (P2MP) forwarding core that connects the edges. The core provides line-speed switching, and a set of network functions that facilitate matching supply with demand and create edge-to-edge paths. Those functions are: Rendezvous (RV), to provide naming, and Topology Management (TM), to provide routing~\cite{Fotiou2012,Fotiou2010}. A programmable core with emerging technologies in Software-Defined Networking (SDN) is advantageous; provided proactive network management that does not introduce control delay. We utilise our solution in~\cite{Reed2016} to facilitate such a SND core. 
An IP edge is connected to the core through a Service Router (SR) that provides mapping of traditional IP-based protocols (IP, HTTP,..ect) to and from a namespace understood by the RV function. Furthermore, the SR  translates the request/receive semantics into an equivalent publish/subscribe counterpart~\cite{Trossen2015}. 
Clients requests are translated into publishing a request for a service, while a server listening for requests is translated into a subscription for requests by the bridging SR. Similarly, a server response is translated by the SR to a publication, while the client waiting for response is translated into a subscription.

The architecture above loosens the tie between information and location, as it treats an IP address, FQDN or URL as services, decoupled from the location of their actuators. Hence matching supply with demand is independent from location, where the latter's awareness is limited to the TM function to create edge-to-edge paths or multicast trees. This renders the DNS service irrelevant and not required to facilitate the communication paradigm. Furthermore, the Pub/Sub model is more aligned with the transport semantics of existing CDNs, which already uses a Pub/Sub model in the transport overlay; particularly that of streaming CDNs~\cite{Kontothanassis2004}.  Forwarding in the core is enabled through fast functions (FW), such as that proposed by~\cite{Jokela2009} and its variant suitable for SDN~\cite{Reed2016}.
We utilise this architecture to propose our novel, flexible CDN (fCDN) solution in the following section. We show that by using the RV for matching supply with demand, we no longer need DNS redirects or DNS extension to provide mapping. Furthermore, the network is natively supporting multicast, thereby eliminating the need for content splitters. 

\section{Flexible CDN Architecture}
\label{sec:cdn_arch}
Based on the initial ideas of IP-over-ICN, described in Section~\ref{sec:ip_over_icn}, the key aspect of formulating our CDN architecture is to interpret individual CDN servers as service endpoints where requests from users to CDNs, as well as from origins to CDNs, are entirely routed at the HTTP service level. The routing for the latter is realized through the ideas presented in the previous Section~\ref{sec:ip_over_icn}. Following this logic, we propose a novel \emph{flexible-CDN} architecture (fCDN), comprized of a set of functions providing labelling, resource monitoring/management and peering capabilities. Our proposal removes the mapping and transport overlays from CDN, and instead provides a simplified Pub/Sub mapping and transport capabilities in the network. This is achieved through a Path Computation Function (PCF), embraced from the ICN proposition described in Section~\ref{sec:ip_over_icn}. We place our solution within the context of a local fCDN in an operator's network, and assume global fCDN to be a connected set of local clusters.

% Here, we propose a novel CDN architecture that has a simplified mapping overlay and no dedicated transport counterpart. Instead, anycast mapping and multicast delivery are supported in the network by use of the ICN functions described in the Section~\ref{sec:ip_over_icn}, namely: RV, TM and FW~\cite{Trossen2015,Jokela2009}. We place our solution within the context of a local CDN in an operator's network and assume global CDN to be a connected set of local clusters.

\begin{figure}[tb]
	\centering
	\subfloat [Functional view]{
				\includegraphics[width=1\linewidth]{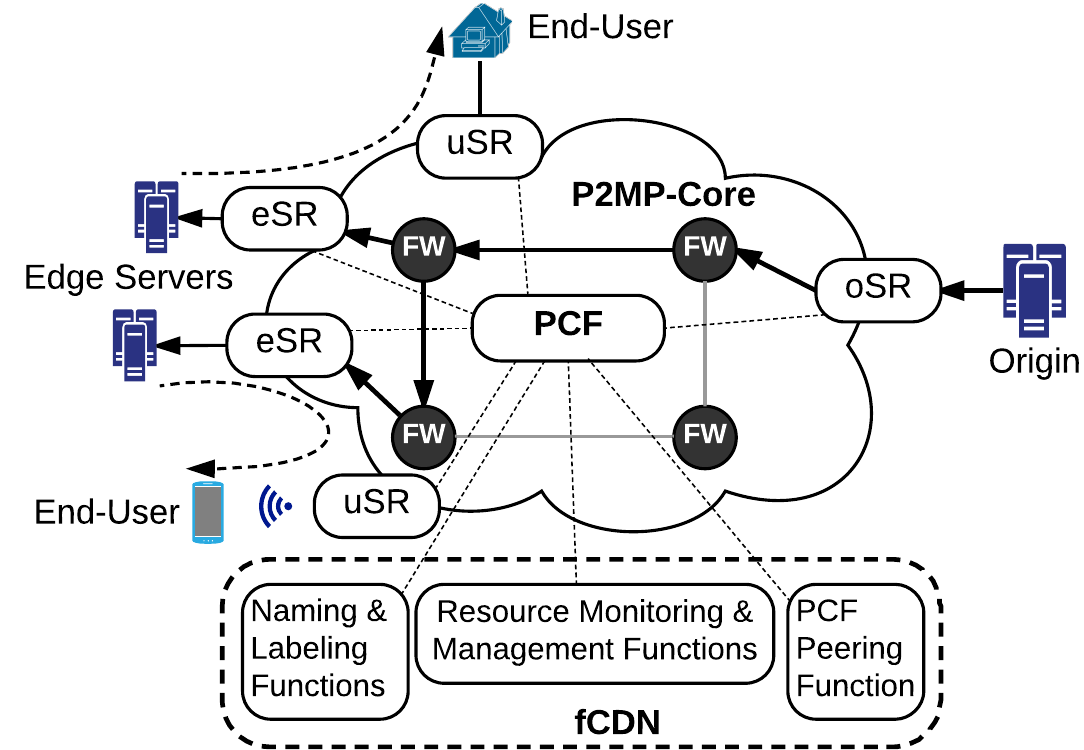}
	}
		\label{fig:CDN_arch}\vfill 
	\subfloat [HTTP/CDN name-space view]
	{
				\includegraphics[width=1\linewidth]{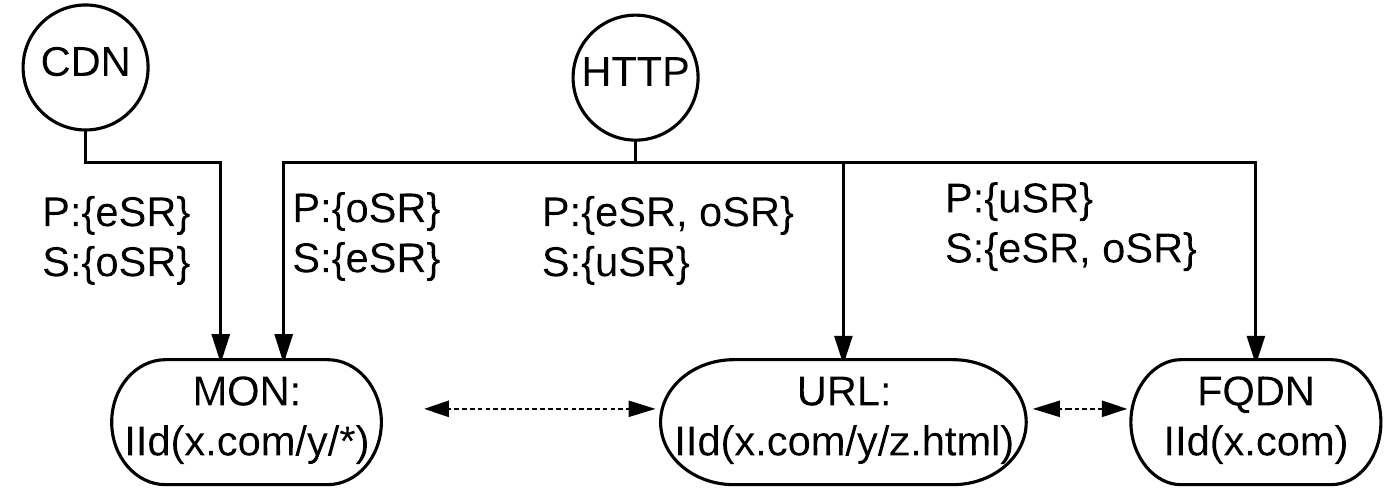}
	}
		\label{fig:CDN_ns}
        \\\vspace{0.5em}
	\caption{A functional and name-space views of our proposed CDN architecture. The functional view, (a), shows the central Path Computation Function (PCF) matching publications with subscription and creating source-routed multicast trees from the origin to edge service points and from edge points to consumption points of end-users}
	\label{fig:CDN}
\end{figure}

The proposed architecture is depicted in Figure~\ref{fig:CDN}(a). A network operator deploys a fast-switching, Point-to-Multi-Point (P2MP) substrate in the core; capable of performing stateless-multicast forwarding. Such a fabric may be facilitated with the Bloom Filter-based solution proposed in~\cite{Jokela2009}, its Bit-based variance in SDN~\cite{Reed2016}, or the Bit Indexed Explicit Replication (BIER) solution~\cite{Giorgetti2017,Wijnands2017}. For the rest of this work, we will assume the solution of~\cite{Reed2016} to be the forwarding mechanism in the network core. 
Edge-to-edge routing in the core is facilitated through the PCF, which matches service requests (i.e., ICN publications) with service availability (i.e., ICN subscriptions), resulting in a suitable path calculation between the matching network elements, as described in~\cite{Fotiou2010,Trossen2015}.
%-----------------------------
%\begin{figure}
%\centering
%\includegraphics[width=\columnwidth]{CDN_arch}
%\caption{Functional view of our proposed CDN architecture showing a central Path Computation Function (PCF), matching publications with subscription and creating source-routed multicast trees from the origin $O$ to edge $E$ service points and from edge points to consumption points of end-users}
%\label{fig:CDN_arch}
%\end{figure}
%-----------------------------
%%-----------------------------
%\begin{figure*}[tb]
%	\centering
%	\begin{subfigure}[t]{0.5\textwidth}
%		\centering
%	\includegraphics[width=\columnwidth]{CDN_arch}
%	\caption{Functional view}
%	\label{fig:CDN_arch}
%	\end{subfigure}%
%	~ 
%	\begin{subfigure}[t]{0.5\textwidth}
%		\centering
%	\includegraphics[width=\columnwidth]{CDN_ns}
%	\caption{HTTP/CDN name-space view}
%	\label{fig:CDN_ns}
%	\end{subfigure}
%	\caption{A functional and name-space views of our proposed CDN architecture showing the HTTP/CDN name spaces with Pub/Sub sets on the right hand side, and the central Path Computation Function (PCF) on the left hand side, matching publications with subscription and creating source-routed multicast trees from the origin to edge service points and from edge points to consumption points of end-users}
%	\label{fig:CDN}
%\end{figure*}
%%-----------------------------
%-----------------------------
%-----------------------------
At the boundary of the core, \emph{SRs} are presented in multiple points; connecting traditional IP-based clusters at the edge over the P2MP-core. The term ``service'' here refers back to the initial idea of IP-over-ICN to interpret any IP-based service, such as HTTP, as a service transaction in the underlying ICN-based network. For this, each SR  provides the essential feature of mapping those services to/from their respective pre-defined Information Identifiers (IIds), grouped under specific name-spaces. For our purposes of content delivery, we extend the existing mapping capability to provide aggregate content mapping and distribution in CDNs. The SR  may have customisable variations to provide different features, depending on the IP edge connected to the function, be it an end-user consumption point or a service point. 

For our CDN purposes, service points can be classified into: \emph{Entry Points} and \emph{Edge Points}. Entry points are nodes that have a stored data copy of all the content that is to be served in the network, for static content; and, have a direct link to content originators, such as broadcast studios, for live content. For the sake of simplicity, without loss of generality, we refer to entry points as \emph{origins} and the SR connecting an origin cluster as oSR. Edge points are service points placed closer to the end-user. They typically have constrained resources and hence only cache a subset of the static content, while fetching the rest (including live content) from origins. We refer to the SR connecting an edge service point as an edge Service Router (eSR).

The fCDN is presented as a set of functions to the network, freed from the DNS mapping and transport complexity of traditional CDN. Instead, the fCDN focuses on: defining service/managed-content names; instrumenting the publish/subscribe behaviour of SRs in the service points, namely eSR and oSR;  monitoring usage behaviour and optimising the utilization of fCDN resources; and, establishing peering relations with other fCDN clusters. We define three fCDN functions to provide these capabilities:\\
\textbf{Naming and Labelling Function (NLF)}: defines the size and aggregation scheme of information objects into aggregate managed objects, `cacheable' in different service points, and provide them with \emph{Managed Object Names (MONs)}. This function is equivalent to the naming function in existing CDN mapping overlay. The function also disseminates service/managed-object names to the eSR and oSR  and provides them with service-based publish/subscribe policies.\\
\textbf{Resource Monitoring and Management Function (RMMF)}: monitors consumption patterns and fCDN load, and, uses this information to manage the fCDN resources; \emph{i.e.} how many service points are active in the network and the set of services/content they serve. This is essential to facilitate dynamic and temporal placement of virtual edge servers in highly dynamic networks.\\
\textbf{PCF Peering Function (PPF)}: manages peering policies, names distribution and path \emph{stitching} between PCFs in multiple Autonomous Systems. Given that inter-domain routing is outside the scope of this work, we will give no further details here; leaving it to our future work.

\subsection{Managed Service and Content Naming}
The name-spaces to be supported in our proposed fCDN may vary for different services. Here we describe a HTTP-based name-space, focusing on HTTP-based services. We show through Figure~\ref{fig:CDN}(b) how the proposed fCDN architecture defines names for managed objects and their respective Pub/Sub relationships.
% These relationships are utilized, as illustrated in Figure~\ref{fig:CDN_msc}, to disseminates managed object from origin service points (\emph{i.e.} origin-oSR instance pair) to edge service points (edge-eSR instance); and, to map and deliver the requested content to end-user. 

\emph{Managed objects} are aggregate bundles of content, named and disseminated as individual objects. Such bundles may be collections of static (VoD) content, or bundles of live streams. Each managed object is given a \emph{Managed Object Name (MON)}; the format of which, is to be defined either by the fCDN provider or by the content provider, through agreement with the fCDN provider. This could be a wild-card URL, a `stream identifier', and/or a service FQDN, as present in existing CDNs~\cite{Kontothanassis2004, Su2008}. Notably, the size of a managed object and the number of objects in a network is a fCDN design parameter, inherited from existing CDNs. This critically differentiates our proposal of naming \emph{managed objects} from existing proposals to naming arbitrary objects in the Global Internet, such as~\cite{Ahlgren2012,Amadeo2016}. Our proposal allows for object naming within well-defined scalability metrics, such as the size of lookup tables and TCAMs entries. Therefore, it is generally expected to incur a much smaller (and manageable) state than that of the name space of arbitrary objects in the Global Internet.    
The HTTP and CDN name-spaces are illustrated in Figure~\ref{fig:CDN}(b), with exemplary FQDN, URL and MON. The `http' root scope is accessible to all uSR, eSR and oSR to deliver HTTP services to end-users; while the 'cdn' root scope is dedicated to eSR and oSR only, and utilized to retrieve managed objects from origins or other edge service points. Details of the Pub/Sub interactions are described in the next section.
\subsection{Publish/Subscribe Model}
A service point, being an origin or an edge, registers a subscription under the `http' root scope in the PCF for requests of all the FQDNs it is designated to serve to end-users. This resembles the action of a server listening for requests. The set of FQDNs to subscribe for is disseminated by the NLF of the fCDN. A service point willing to provide a cached object (a.k.a managed-object) to other service points also subscribes to the MON under the `cdn' root scope. This resembles listening to requests of MONs, issued by other service points. When a provider point needs to respond to a request it publishes the same MON under the `http' name-space. This publication is the response to the original request that was published under the `cdn' root scope.
%-------------------------------------------------------------------------
\subsection{End-to-End Content Distribution}
\begin{figure}
\centering
\includegraphics[width=\columnwidth]{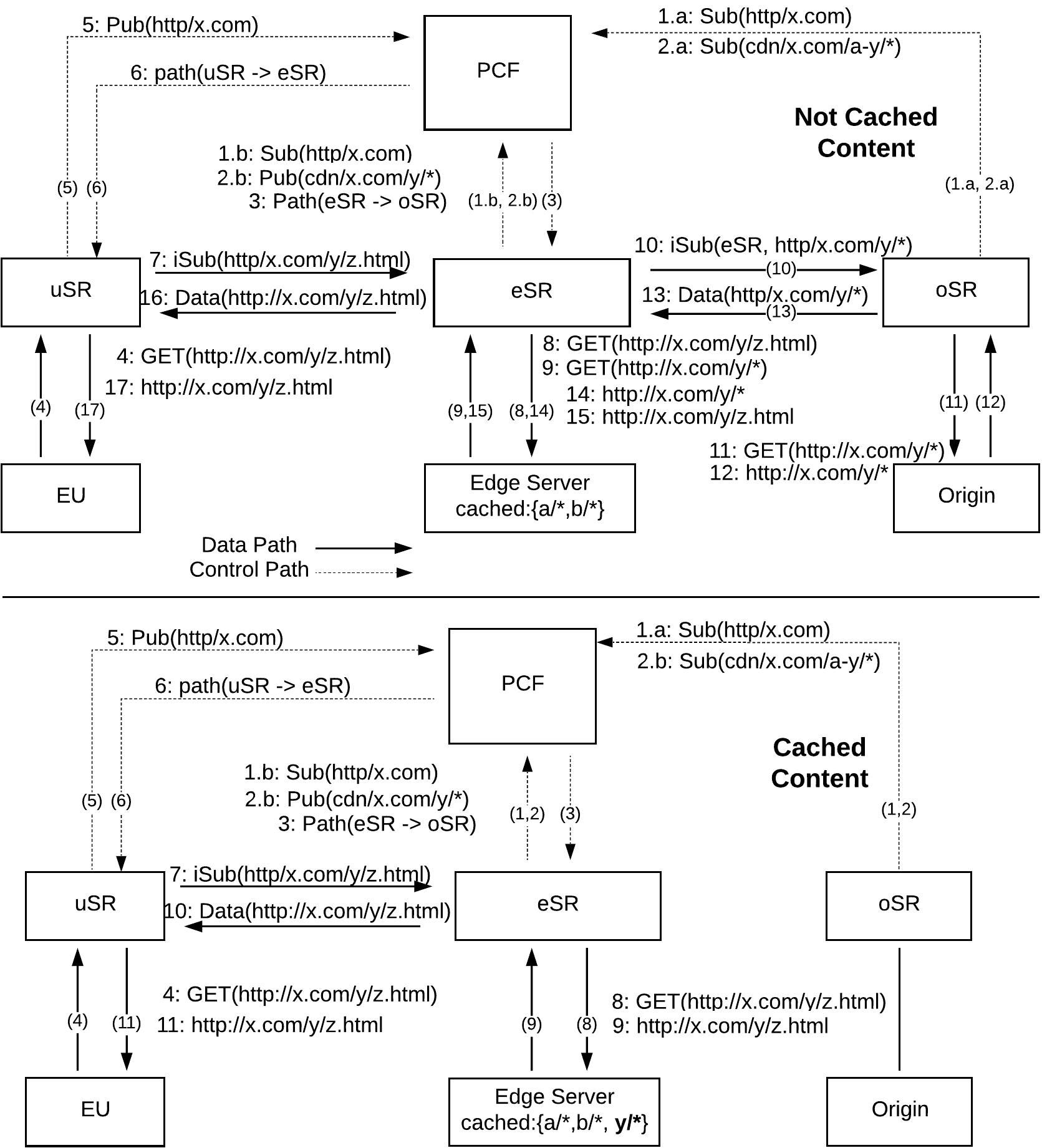}
\caption{Detailed overview of exchanges between an End-User (EU) and surrogate, and between a surrogate and origin to facilitate content delivery to the customer}
\label{fig:CDN_msc}
\end{figure}

We describe end-to-end content distribution through a video streaming example by the mobile end-user in the network of Figure~\ref{fig:CDN}(a), both when content is already cached and when it is not. Figure~\ref{fig:CDN_msc} illustrates the sequence of messages exchanged between the different network functions to facilitate content delivery in this scenario. When the fCDN is started and updated, the NLF  would determine the set of FQDNs/MONs to be serviced in the network, depending on the input provided by content providers. Meanwhile, the RMMF defines the number and location of service points in the network, as will be described in Section~\ref{sec:model_and_problem}, and the set of FQDNs/MONs to be serviced by each point. Communication between the two functions results in a list of FQDNs/MONs to be distributed to the eSR and the oSR, by the NFL.  Both the eSR and the oSR subscribes to the list of FQDNs, while only the oSR subscribe to 'cdn/MON' and only the eSR publish `cdn/MON'. The PCF detects a match between a publication and a subscription, and provides the publisher eSR with a path to reach its `best' oSR. To this end, the path is cached in the eSR, but the content is not necessarily requested.

A mobile end-user wishing to watch a HTTP video, would send a HTTP request request to its uSR. The latter maps the FQDN of the request to an information Id and publishes a FQDN request to the PCF, under the `http' name-space. The PCF would then match the publication with all subscriptions from eSRs and oSRs that match the FQDN, and establish a delivery path between the uSR  and the 'best' subscriber SR. In this case the best SR is the eSR closest to the end-user. The PCF supplies the path to the uSR, which uses it to forward the HTTP request to the eSR. The eSR extracts the HTTP request and passes it to its paired edge server. If the edge server has the information cached, it would respond back with the data through its eSR to the uSR and into the end-user. Contrarily, if the server does not have a cached copy, it will request the content from another server. The eSR, maps the request into a `cdn/MON' publication. The PCF matches the `cdn/MON' publication with existing subscription(s) to the same name and creates a delivery path from the eSR to the oSR. The PCF provides the eSR with the path, which is then used to send the MON request to the oSR. After which, the eSR subscribes for the oSR response under the `http/MON' path. Meanwhile, the oSR receives the request and pass it to the origin server. The origin server responds back, through the oSR, with the managed-object identified by the MON.
Once the edge server receives the content, it will respond to the end-user request with the video content. Figure~\ref{fig:CDN_msc} provides a detailed overview of these interactions.

When the content is VoD, the serving point may either be an origin or an edge, with a cached copy of the video; whereas, if the content is a live stream, the service point is the content originator. In the latter case, the origin is provided with a direct path to the edge service point, without the need for an intermediate reflector.
Notably, the above operations are only required on the first packet, even in a chunk-based HTTP stream; whereby, requests for subsequent chunks are forwarded directly to the eSR. Using a source-routed forwarding solution such as that described in~\cite{Jokela2009,Reed2016}, both the origin-edge, and edge-edge streaming can be multicast, as it has been described in the previous Section~\ref{sec:u2m2u}.

% The RM overlay is responsible of optimising the number of origins and surrogates in the network, and the set of FQDNs and MONs that they should serve. 
% Recall that origins here refer to \emph{entry points}, the CDN interface to the actual content originators, such as production studios. 

% Here, we propose a novel CDN architecture with a considerably simplified mapping, transport and routing overlays, while preserving the anycast and multicast attributes of CDNs. Our proposal is placed within the context of a single operator's network, having intra-domain considerations. Global CDN architectures is assumed to be a connected set of local CDNs, and hence the proposal is applicable with suitable extensions to facilitate peering relations. Details of inter-domain extensions are outside the scope of this work.

% The proposed architecture assumes an IP-based unicast network at the edge and a point-to-point network in the core, having source-routed multicast capabilities, similar to that of~\cite{Jokela2009}. Routing in the core is provided through an Information-Centric Networking (ICN) function, namely the Path Computation Element (PCE). The latter consists of two parts: an RV function to match supply with demand and a TM function to create delivery paths/trees.

%-----------------------------
%-----------------------------
%-------------------------------------------------------------------------
\subsection{HTTP Unicast-to-Multicast-to-Unicast}
\label{sec:u2m2u}
Our architecture has a key advantage over existing HTTP streaming mechanisms~\cite{Kontothanassis2004, Sitaraman2014}, that is the ability to spontaneously multicast the response back from the server to synchronous or quasi-synchronous clients, be them end-users or edge service points. This advantage is critical in providing scalable content dissemination over the network. It provides a more flexible and dynamic substitution to the transport overlay of existing CDNs, using a network of \emph{reflectors}. The improvement in flexibility and dynamicity in our solution stems from the ability to change the shape and size of the multicast tree, without having to adhere to pre-defined `splitting' points; and, therefore, not having to address bottleneck challenges on the 'splitting' points. Moreover, unlike existing CDNs~\cite{Kontothanassis2004, Mukerjee2015}, our solution not only supports multicast between the origin and the edge service point; but, also between a service point and user consumption point (i.e. end-users uSR pair). For a large number of users, our solution allows for scalable delivery of HTTP services using substantially less network capacity. 

Following Figure~\ref{fig:CDN_msc}, when a group of quasi-synchronous end-users request new content (e.g. HTTP chunk), the uSR would only issue one publication to the PCF. The uSR then forwards a subscription to the HTTP response to the eSR, on behalf of each of the end-users. If these subscriptions fall within a pre-defined period of time, namely the \emph{catchment interval}, the eSR would formulate a multicast group of one uSR and issue a single response back, using the unicast forwarding identifier - from eSR to uSR - provided by the PCF. If, supposedly, another uSR issues a subscription to the eSR response within the same catchment interval, then in this case, the eSR extends the multicast group to two uSRs and uses the two unicast forwarding identifiers to create a single multicast identifier, following the mechanism of~\cite{Jokela2009} or~\cite{Reed2016}. The latter is used to multicast the response to the uSR.

Notably, in both of the cases above, the eSR passes only the first request to the server, while suppressing subsequent requests within the catchment interval; thereby, the server receives a single request; for which, it issues a single response. Intuitively, this reduces the load on the server by the number of requests falling within the catchment period.
Similarly, when two or more eSRs subscribe to the content of the same MON within the catchment interval of the oSR, the latter would create a multicast group of the eSRs and publish a single, multicast, response back to them all.
Moreover, while end-users are quasi-synchronous, edge service points can be orchestrated to subscribe for MON content within a predefined period; thereby, engineering the size of multicast groups according to the target capacity saving. Our analytical formulation in Section~\ref{sec:model_and_problem} provides a formal definition of the average size and number of multicast groups and the relationship between that and the catchment interval and the total transmission period of content.

In an ideal scenario, content would be cached in every node. However, given constrained link, storage and processing capacities in current and future networks, optimized resource utilization and service delivery is critical to enable the expected growth in 5+G networks. Similar to~\cite{Wichtlhuber2015}, we assume a fCDN/operator collaboration to provide an accurate optimization of resources by the RMMF, depending on the conditions of the CDN and the operator's networks. Our collaboration comes in the form of exchanged CDN/network statistics between the RMMF and the PCF.
Next we model our proposed CDN network and formulate the optimization problem facing the resource management overlay.

%-------------------------------------------------------------------------

\section{Model and Problem Formulation}
\label{sec:model_and_problem}
Here, we model our proposed fCDN architecture and formulate the optimization problem of dimensioning network and storage capacities. Notice that in our fCDN architecture, the publish/subscribe paradigm is carried out in both the request and response directions, with appropriate publish/subscriber roles as required by the nature of an IP transmission (publication) or reception (subscription). In the direction of requests from clients, service points act as subscribers for the request information; in contrast in the response direction service points act as publishers of the content associated with an FQDN or a MON. For the sake of simplicity and clarity, we will describe our model in relation to the response direction, where a service point assumes the role of a content \emph{publisher} and the end-user or a non-cached edge service point takes the role of a content \emph{subscriber} (hiding the fact that for bi-directional TCP/IP semantics the service point also acts as a \emph{subscriber} when listening and the client \emph{publishes} the first connection packet \emph{etc.}).

\subsection{Modelling Traffic}
\label{sec:model}
First, we model the overall collection of FQDNs and MONs offered by multiple content providers as a global set of ranked items 
% $I = \{\langle i_a, r_a, b_a, c_a \rangle | \ a \in \{1, 2,...,n,...,N\}, \ r_a, b_a, c_a \in \R_+\}, \ |I| = N$;
$I = \{\langle i, \varphi_i, \eta_i, \zeta_i \rangle | , \ \epsilon_i, \eta_i, \zeta_i \in \R_+\}, \ |I| = N$; 
where $i$ is the information Id of the item, $\varphi_i$ is the probability of occurrence of $i$, $\eta_i$ is the network capacity required to deliver $i$ and $\zeta_i$ is the storage requirements to store/cache $i$. An information item $i$ may be grouped under one of two name-spaces: either just the \emph{HTTP name-space} or both the \emph{HTTP name-space} and the \emph{CDN name-space}, as shown in Fig~\ref{fig:CDN_ns}. Recall that the HTTP namespace is utilized for all HTTP communication between uSR and the e/oSR, whereas the CDN name-space is only used between eSR and oSR to fetch managed content, through their MON. 
% The publish/subscribe paradigm is carried out at both the request and response directions, with roles of publishing/subscribing revered in each direction. For the sake of simplicity, we will focus our model on the response direction, where a service point assumes the role of a response \emph{publisher} and the end-user or a non-cached edge service point takes the role of a response \emph{subscriber}.

An operator's network is modelled as a directed graph $G = \{V, A\}$; having a set of vertices $V$, connected by a set of arcs $A = \{\langle u,v \rangle | \ u , v \in V , \ u \neq v\}$.
% Each edge $<u,v>$ has a link capacity, defined as $b_{<u,v>} = \min(u,v)$. 
Each $v \in V$ has a pre-defined storage and network capacities. The storage capacity is constrained by the amount of storage resources a CDN provider is willing to place on site of $v$.
% Storage capacity is constrained by the size and number of hard discs a CDN provider is willing to place in one location. 
It can be defined as $C = \{c_v \in \R_+| \ \forall v \in V\}$. The network capacity of a vertex, $v \in V$, is here defined as the total network capacity on all outgoing edges from the vertex to its neighbours where the link capacity on an outgoing edge is $b_{\langle u,v \rangle}$. We define the neighbourhood of $u \in V$ as a subset $V_u = \{ v | \ v\in V, \ v \neq u,\; \langle u,v \rangle \in A\}$.
Thus, we define network capacity that vertex $u$ can provide as 
\begin{equation}
B_u = \sum_{v \in V_u} b_{\langle u,v \rangle}, \ \forall u \in V
\end{equation}
and the overall network capacity, across all network links, as
\begin{equation}
B = \sum_{\langle u,v \rangle \in A} b_{\langle u,v \rangle}
\end{equation}

% In our CDN architecture, the publish/subscribe paradigm is carried out at both the request and response directions, with roles of publishing/subscribing revered in each direction. For the sake of simplicity, we will focus our model on the response direction, where a service point assumes the role of a response \emph{publisher} and the end-user or a non-cached edge service point takes the role of a response \emph{subscriber}. \mathcal{O}
We assume a CDN provider plans to place in the network a set of origins $O \subseteq V$ and a set of edge service points $E \subseteq V,$ where $\ E \cap O = \emptyset$.  
Origins, $O$, are large-scale data centers provided with storage and network capacities large enough to accommodate the content of all the items in the catalogue $I$. In contrast, edge service points, $E$, are nodes that are closer to the end-user and have tighter constraints on storage and network capacities than origins. The SR of all these service points behave as \emph{publishers}, $P = E \cup O$, of the content associated with the items in $I$; regardless of whether or not the content is actually stored in the server associated with the edge service point.
Hence, edge points, specifically the eSR, may advertise the availability of an FQDN but they will not necessarily cache all the content associated with the FQDN. Nonetheless, provided with the information Ids of pre-defined MONs of the FQDN, an eSR may subscribe proactively to a MON, in the CDN name-space, to obtain and cache paths to the origins of the MON. 

The network presents content \emph{subscribers}, $S \subseteq V$, that are SR that are interested in advertized information. Subscribers may either be uSR, fetching content on behalf of their connected users; or, eSR, retrieving content for their edge servers that is not cached. A particular item, $i$, will have certain subscribers $S_i \in S$ that can be obtained from appropriate publishers $P_i \in P$ that have either cached the item or have a path to it.
% A subscriber subscribes to a subset of items $I_s \subseteq I$ in the PCF; and, a single item may be subscribed to by a set of nodes $S_i = \{s \mid i \in I_s, s \in S \}$. 
Notably, a single node can be both a publisher and subscriber for content. For instance a node connecting a set of end-users as well as an edge server to one (u/e)SR, belongs to both $S$ and $P$. Thus often $S_i \cap P_i \neq \emptyset$. In fact, in an ideal scenario, where all content items are cached locally, $S_i \equiv P_i, \forall i \in I$. However, given the significant operational and technical costs for content placement and distribution, often only a subset of items is advertized and/or cached, while the rest is delivered through the network from origins.
The set, $D$, describes publishers and subscribers mutually interested in information for the items $I$ and is 
$D  = \{\langle i, P_i, S_i \rangle | S_i \subseteq S, \ P_i \subseteq P, \ \forall i \in I\}$

For each subscriber $s \in S_i$ of the $ith$ tuple in $D$, the PCF creates a Pub/Sub relationship, $\lambda_{\langle i, s, p \rangle}$ with a publisher $p \in P_i$, defined as
\begin{eqnarray}
\lambda_{\langle i, s, p \rangle} = \begin{cases}
        			               1, \ \textrm{if } p \textrm{ selected to serve } i \textrm{ to } s \\
					               0, \ \textrm{otherwise}
									\end{cases}
\end{eqnarray}
The overall set of Pub/Sub relationships of the $ith$ tuple, $\Lambda_i$, is defined as the set of all Pub/Sub relationships which match each subscribers of $i$ to a publisher of $i$ noting that for a set of subscribers for $i$ there may be multiple publishers such that 
\begin{equation}
\Lambda_i  = \{\lambda_{\langle i, s, p \rangle}\ | \ i \in I, \ p \in P_i, s \in S_i\} \label{eq:i_pub/subs}
\end{equation}
and for all items
$\Lambda=\{\Lambda_i\ | i \in I\}$

The challenge here is to determine:
\begin{enumerate*}
\item how many publishers $P$ to place in the network
\item how to split $P$ between $E$ and $O$, and
\item where to place $O$ and $E$, such that the maximum distance to any $s \in S$ is kept to a minimum.
\end{enumerate*}
The first challenge can be addressed by reducing the problem to a variance of the ``$K$-center'' placement problem~\cite{Potikas2009}. This problem is known to be NP-hard~\cite{Potikas2009, Hillmann2016} such that an optimum solution cannot be found in polynomial time. Hence we will address it through greedy algorithms and show that they provide a reasonably good and fast solution.
%-------------------------------------------------------

\subsection{Storage and Network Requirements}
\label{sec:requirements}
To address the placement problem, we first model the storage and network requirements of the items with respect to both the publishers and the network. We define the choice of storage of an item $i \in I$, with storage requirement $\zeta_i$, at a publisher $p \in P$, as
\begin{eqnarray*}
\alpha_{i,p} & = & \begin{cases}
					     \zeta_i, \ \textrm{if } i \textrm{ stored in } p \\
					     0, \ \textrm{otherwise}
				\end{cases} \label{eq:storage_def} \\
\end{eqnarray*}
This means, the storage requirement of a publisher $p \in P$ can be determined as:
\begin{eqnarray}
 \theta(p) & = & \sum_{i \in I} \alpha_{i,p} \label{eq:storage_p_def}\\
\text{s.t.} \ \ \theta(p) & \leq & c_p, \ \forall p \in P \label{eq:storage_cons}
\end{eqnarray}
where the constraint of~\eqref{eq:storage_cons} ensures that the total storage demand of all item published by $p$ does not exceed the storage capacity of $p$.
% Moreover, based on the storage definition of~\eqref{eq:storage_def}, the total storage footprint of item $i$ can be defined as:
% \begin{equation}
% \alpha_{i} = \sum_{p \in P_i} \alpha_{i,p} \label{eq:ps_storage_def}
% \end{equation}

% Now, we define the network footprint of $i$ in a single flow on an arch $\langle u,v \rangle$ as \mjreed{come back to use of footprint}:
% \begin{eqnarray}
% \beta^{i}_{\langle u, v \rangle} & = &  \begin{cases}
% 					     \eta_i, \ \textrm{if } i \textrm{ transmitted over } \langle u,v\rangle \\
% 					     0, \ \textrm{otherwise}
% 				\end{cases}
% \end{eqnarray}

% The network capacity required to distribute $i$ is dependent on the dissemination mode applied to satisfy $\Lambda_i$ set of relationships. That, in turn, is dependent on the semantics of the edge services connected through our architecture. Recall here that each $s \in S$ is either a uSR instance or eSR instance connecting one or more IP-based (HTTP) clusters. We define $M_s$ as the total number of HTTP clients connected to $s \in S$; and, $m_{s,i} \leq M_s$ as the number of HTTP clients requesting/receiving item $i$. 
% Given the unicast nature of the HTTP semantics, we assume initially that the HTTP clients are provided through individual, parallel, unicast micro-flows. This means that for $m_{s,i}$ number of clients of $i$ behind $s$, the capacity required on arch $\langle u,v \rangle$ to satisfy the Pub/Sub relationship $\lambda_{\langle i,p,s \rangle}$ is defined as:
The network capacity required to distribute $i$ is dependent on:
\begin{enumerate*}
\item the subset of publishers selected to deliver to the subscribers, and
\item the number of streams established to deliver the content to the subscribers.
\end{enumerate*}
To identify the number of selected publishers, we go back to Pub/Sub relationships $\Lambda_i$ of~\eqref{eq:i_pub/subs}. We define the set of Pub/Sub relationships of $i$ satisfied by $p \in P_i$ as:
\begin{eqnarray}
% \Lambda_{i,p}  =  \{\lambda_{\langle i,p,s \rangle} | p \in P, \lambda_{\langle i,p,s \rangle} \neq 0 , \ \forall s \in S_{i,p}\} \ , \; \; \Lambda_{i,p}  \subseteq  \Lambda_i\label{eq:i_p_pub/sub}
\Lambda_{i,p} & = &  \{\lambda_{\langle i,p,s \rangle} | p \in P, \lambda_{\langle i,p,s \rangle} \neq 0 , \ \forall s \in S_{i,p}\} \; \; \; \; \label{eq:i_p_pub/sub}
\\
\Lambda_{i,p} & \subseteq & \Lambda_i \nonumber
\end{eqnarray}
These relationships are the basis to calculate the expected load on the server.
Next, we need to define the number of streams to be issued to $S_{i,p}$ by $p$. Recall here that each $s \in S$ is either performing uSR or eSR, connecting one or more IP-based (HTTP) clusters. We define $M_s$ as the total number of HTTP clients connected to a node $s \in S$; and, $m_{s,i} \leq M_s$ as the number of HTTP clients requesting/receiving item $i$. 
Given the unicast nature of the HTTP semantics, we assume initially that the HTTP clients are provided through individual, parallel, unicast streams, ignoring the multicast capability of the architecture until later. Consequently, for each Pub/Sub relationship $\lambda_{\langle i,p,s \rangle} \in \Lambda_{i,p}$, where $s$ has $m_{s,i}$ clients, the capacity required on network link $\langle u,v \rangle$ to satisfy $\lambda_{\langle i,p,s \rangle}$ is 
\begin{eqnarray}
\beta^{\lambda_{\langle i,p,s \rangle}}_{\langle u, v \rangle} & = & \begin{cases}
					               \eta_i m_{s,i}, \ \textrm{if } \langle u, v \rangle \textrm{ used for } \lambda_{\langle i,p,s \rangle} \; \; \\
					               0, \ \textrm{otherwise}
									\end{cases} \label{eq:link_cap_def}
\end{eqnarray}
For publisher $p$, the network outgoing capacity required to deliver all Pub/Sub relationships of $i$ is:
\begin{eqnarray}
\chi(i,p) & = & \sum_{v \in V_p}  \sum_{\lambda \in \Lambda_{i,p}} \beta^{\lambda}_{\langle p, v \rangle}
\end{eqnarray}
where the subscripts on $\lambda_{\langle i,p,s \rangle}$ have been dropped for brevity.
Recall that $V_p \subseteq V$ is the neighbourhood set of $p$. Now, the total outgoing capacity to deliver all tuples of $\Lambda$ is
\begin{eqnarray}
\chi(p) & = & \sum_{v \in V_p} \sum_{\Lambda_{i,p} \in \Lambda} \sum_{\lambda \in \Lambda_{i,p}} \beta^{\lambda}_{\langle p, v \rangle} \\
\text{s.t} \ \ \chi({p}) & \leq & B_p, \ \forall p \in P \label{eq:net_cons}
\end{eqnarray}
where constraint \eqref{eq:net_cons} ensures that the capacity demanded from a publisher does not exceed its network capacity.

For any $\langle u,v \rangle \in A$, the capacity required to deliver all Pub/Sub relationships of all tuples of $\Lambda$ is
\begin{eqnarray}
\chi(\langle u, v \rangle) & = &  \sum_{\Lambda_{i} \in \Lambda} \sum_{\lambda \in \Lambda_{i}} \beta^{\lambda}_{\langle u, v \rangle}\label{eq:ps_link_def} \\
\text{s.t} \ \ \chi({\langle u,v \rangle}) & \leq &  b_{\langle u, v \rangle}\label{eq:ps_link_cons}
\end{eqnarray}

% Now, we calculate the total network capacity required on $\langle u,v \rangle$ as:

% This means the total capacity required on $\langle u,v \rangle$ is:
% % \begin{eqnarray}
% % \sum_{i \in I} \sum_{s \in S_i} \beta^{\lambda_{\langle i,p,s \rangle}}_{\langle u, v \rangle} & \leq &  b_{\langle u, v \rangle} , \ \forall \langle u,v \rangle \in A \label{eq:ps_link_cons}\label{eq:ps_link_def}
% % \end{eqnarray}

% \begin{eqnarray}
% \chi({\langle u,v \rangle}) & = & \sum_{i \in I} \sum_{s \in S_i} \beta^{\lambda_{\langle i,p,s \rangle}}_{\langle u, v \rangle}\label{eq:ps_link_def} \\
% \text{s.t} \ \ \chi({\langle u,v \rangle}) & \leq &  b_{\langle u, v \rangle}\label{eq:ps_link_cons}
% \end{eqnarray}
% The constraint of~\eqref{eq:ps_link_cons} ensures that the network capacity requirement does not exceed the link capacity.

% Following the definitions of~\eqref{eq:link_cap_def} and~\eqref{eq:ps_link_def}, the network outgoing capacity required at publisher $p$ is defined as:
% \begin{eqnarray}
% \sum_{v \in V_p} \sum_{i \in I} \sum_{s \in S_i} \beta^{\lambda_{\langle i,p,s \rangle}}_{\langle p, v \rangle} & \leq & B_p, \ \forall p \in P \label{eq:net_cons}
% \end{eqnarray}

% \begin{eqnarray}
% \chi(p) & = & \sum_{v \in V_p} \beta_{i, \langle p, v \rangle} \\
% \text{s.t} \ \ \chi({p}) & \leq & B_p, \ \forall p \in P \label{eq:net_cons}
% \end{eqnarray}
% Where $V_p \subseteq V$ is the neighbourhood set of $p$. The constraint of~\eqref{eq:net_cons} ensures that the capacity demanded from a publisher does not exceed its network capacity.
%-------------------------------------------------------
\subsection{Network Requirements: Multicast}
\label{sec:mc_req}
The definition of required link capacity in~\eqref{eq:link_cap_def} is applicable to existing HTTP streaming mechanisms, both for static (VoD) and live content. However, our solution provides the ability to multicast the response back to a group of clients, if their requests happen to be synchronous or quasi-synchronous within a predefined catchment interval, as described in Section~\ref{sec:u2m2u}.

To define the size of multicast group, and the number of multicast groups, let us first define $T$ to be the total time period required to stream a video item $i \in I$.
$T$ is a relatively short time period, in the order of $0-2$ hours; during which, the popularity of $i$ is invariant~\cite{Traverso2013}. Let us also define $\tau < T$ as the catchment interval. The questions now are: 
\begin{enumerate*}
\item how many users fall within a catchment period (i.e. size of a group), and
\item how many catchment intervals occur during $T$ (i.e. number of groups formed).
\end{enumerate*}
In an orchestrated synchronization of subscriptions, these can be engineered by the service orchestrator, e.g. CDN RMMF. 
However, when client subscriptions are only quasi-synchronous, the catchment interval can be triggered to start by the arrival of the first subscription to the serving SR. Here, the size and number of multicast groups is directly related to the request rate. 

When delivering an item $i$ to $S_i$, multicast can be formed in two ways:
\begin{enumerate*}
\item multicast to multiple clients behind a single subscriber $s \in S_i$, and
\item multicast to a subset of subscribers of $S_i$.
\end{enumerate*}
and for popular items both mechanisms will be be present.
To define the size of a multicast group triggered by a single $s \in S_i$, let $m^t_{s,i}$ be an independent number of HTTP clients subscribing for $i$ at time $t$ and triggering the start of a $\tau$ catchment interval. Recall in Section~\ref{sec:requirements}, we defined $m_{s,i}$ as the number of clients connected to $s \in S$ and actively subscribing for $i$, thus we define $m^t_{s,i}$ as the number subscribing time $t$. Now, the size of the multicast group triggered by $s \in S$  during a $\tau$ duration interval at $p$ is:
\begin{equation}
\gamma^t_{p,i} = \sum_{j = t}^{t + \tau} m^j_{s,i} \label{eq:mc_s_time_dependent}
\end{equation}
The definition of~\eqref{eq:mc_s_time_dependent} renders the multicast group size a time-dependent function.

To estimate the average multicast group size, we refer to the average subscription rate.
Since the client subscriptions are independent from each other, the number of subscriptions per catchment interval is assumed to have a Poisson distribution. Moreover, since the popularity of the item is invariant across $T$, the Poisson distribution is assumed homogeneous, as confirmed by the findings of~\cite{Traverso2013}. Hence, the arrival rate of requests from $s$ for item $i$ as $\mu_{s,i}$ (inter-arrival rate $1/\mu_{s,i}$) is
$\mu_{s,i} = m_{s,i}/T$
and the average group size is
$\mean \gamma_{p,i} = \mu_{s, i}  \tau$.

So far, we have defined the size of a multicast group triggered by a single $s \in S_i$. This translates into a single multicast stream to deliver the content  as opposite to an average number of unicast streams $\mu_{s,i}$ per interval $\tau$. Now, to define the size of a multicast group that spans a subset of subscribers of $S_i$, we refer back to the Pub/Sub relationship $\Lambda_{i,p}$ of~\eqref{eq:i_p_pub/sub}.
% we go back to $\Lambda_i$ set of Pub/Sub relationships of $i$. We define $\Lambda_{i,p} = \{\lambda_{\langle i,p,S_{i,p} \rangle} | S_{i,p} \subseteq S_i, \ p \in P, \ \lambda_{\langle i,p,S_{i,p} \rangle} \neq 0 \}, \ \Lambda_{i,p} \subseteq \Lambda_i$ as the set of Pub/Sub relationships of $i$ satisfied by $p$. 
This set of relationships results in creating a single source multicast tree, with multicast groups triggered by each $s \in S_{i,p}$. The question here is: how many of these groups are synchronous with each other,  as those can be merged into a single multicast tree from the publisher $p$. To answer this question, we consider the total number of clients actively independently subscribing for $i$ during the time period $T$. For a large number of clients, the exponentially distributed inter-arrival time between events is small; and it gets smaller as the number of clients increase. This translates into increasing likelihood of triggering catchment intervals with smaller time gaps between them. Accordingly, the average group size is
% \begin{equation}
% \mean \gamma_{p,i} = \sum_{s \in S_{i,p}} \mu_{s, i}  \tau
% \end{equation}
% \mjreed{I think you want}
\begin{equation}
\lim_{m_{s,i} \to \infty} \ \ \mean \gamma_{p,i} = \sum_{s \in S_{i,p}} \mu_{s, i}  \tau
\end{equation}
This means that in a particular interval of length $\tau$, a single multicast stream is used to deliver the content of $i$  as opposite to an average $\sum_{s \in S_{i,p}} m_{s, i}$ unicast streams.

Next, we determine the number of catchment intervals triggered at $p$ for the entire period of the content $T$. This translates into number of multicast streams that need to be accounted for in the network. First, let $\omega_{s,i}$ be the number of catchment intervals triggered by $s \in S_i$  within the Pub/Sub relationship $\lambda_{\langle i, p, s \rangle}$ and during time period $T$. This means that on average $s$ divides $T$ into intervals :
\begin{eqnarray}
\label{eq:TotalTime}
T & = & \mean \omega_{s,i} \tau + \frac{\mean \omega_{s,i} -1}{\mu_{s,i}}
\end{eqnarray}

Using \eqref{eq:TotalTime} we can now describe the average number of catchment intervals, $\mean \omega_{s,i}$, as a function of $T$:
\begin{equation}
\mean \omega_{s,i} = \frac{T + T/\mu_{s,i}}{\tau + T/\mu_{s,i}} \label{eq:mc_number}
\end{equation}

% This means a single $\lambda_{\langle i,p,s \rangle}$ would require on average $\mean \omega_{s,i}$ number of multicast streams to deliver the content of $i$.
Now, for $\Lambda_{i,p}$ set of relationships originating from $p$, each $s$ would be generating on average $\mean \omega_{s,i} , \ \forall s \in S_i$. For a small number of clients, the likelihood of synchronized triggering of $w_{s,i}$ is small. However,  as the number of clients in the network increases, the likelihood of having synchronized triggering of catchment intervals increases, as the weight of the term $T/\mu_{s,i}$ in~\eqref{eq:mc_number} decreases, rendering the average number of multicast groups a function of $T$ and $\tau$. In this case the average number of multicast streams triggered at $p$ is:
\begin{equation}
\label{eq:mc_T_over_t}
\mean \omega_{p,i} \approx T/\tau
\end{equation}
Now, assuming \eqref{eq:mc_T_over_t}, we can obtain the mean network capacity requirements on a network link $\langle u,v \rangle$ to satisfy the Pub/Sub relationships $\Lambda_{i, p}$ as
%\begin{eqnarray}
%\beta^{\Lambda_{i,p}}_{\langle u, v \rangle} & \approx & \begin{cases}%
%					               \eta_i \omega_{s,i}, \ \textrm{ if }v = s \\
%                                   \eta_i  \omega_{p,i}, \textrm{ if } u = p \\
%                                   [\eta_i \omega_{s,i}, \eta_i \omega_{p,i}] \textrm{ if } u,v \neq s,p\\
%					               0, \ \textrm{otherwise}
%									\end{cases} \label{eq:mc_link_cap_def}
%\end{eqnarray}
\begin{eqnarray}
\beta^{\Lambda_{i,p}}_{\langle u, v \rangle} & \approx & \begin{cases}
                                   \eta_i  \omega_{p,i}, \textrm{ if } \langle u, v \rangle \textrm{ is on the tree}\\
					               0, \ \textrm{otherwise}
									\end{cases} \label{eq:mc_link_cap_def}
\end{eqnarray}
And, the total capacity requirement on $\langle u,v \rangle$ as
\begin{equation}
\label{eq:total_link_traffic}
\chi({\langle u,v \rangle} ) = \sum_{\Lambda_i \in \Lambda} \sum_{\lambda \in \Lambda_i} \beta^{\Lambda_{i,p}}_{\langle u, v \rangle}
\end{equation}
Clearly~\eqref{eq:total_link_traffic} is highly dependent on the choice of publisher, paths and underlying network topology.
%-------------------------------------------------------
\subsection{Problem Statement}

Given the network and communication model presented in Section~\ref{sec:model}; and, the storage and network capacity requirements, defined in Sections~\ref{sec:requirements} and \ref{sec:mc_req}, we define the network dimensioning problem as a multi-objective, weighted, minimization variance of the ``K-center'' problem:\\
\textbf{Definition:}\\
Let $K = |P|$ is the number of service points a CDN provider is planning to place in a network.\\
\textbf{Goal:}\\
Find a set of origin points $O \subseteq P$ and a set of edge points $E \subseteq P$, such that $|E| + |O| = K$, and:
\begin{eqnarray}
\label{eq:optm_prob_def}
% \min: \theta(p), \chi(p), \chi({\langle u,v \rangle}) , \ \forall p \in P , \langle u,v \rangle \in A
\min\limits_{\theta(p), \chi(p), \chi({\langle u,v \rangle})}(\theta(p) + \chi(p) + \chi({\langle u,v \rangle})) , \ \\ \forall p \in P , \langle u,v \rangle \in A \nonumber
\end{eqnarray}
\textbf{Subject to:}
the storage capacity constraint of~\eqref{eq:storage_cons};
the link capacity constraint of~\eqref{eq:ps_link_cons};
the network capacity constraint of~\eqref{eq:net_cons}; and,
traditional flow conservation constraints%~\cite{Ahuja1993}

Next, we describe our proposed greedy heuristic algorithm to solve this problem. We compare it later on in our Evaluations~\ref{sec:eva} against reference heuristics based on node \emph{closeness} centrality or population size.

\section{Algorithmic Solutions}
\label{sec:algo}
The ``K-center'' problem is found to be NP-hard; that is, an optimal solution cannot be found in polynomial time. Consequently, here we propose a fast greedy algorithm to find a near-optimum solution of good quality and within small period of time. We first relax the constraint of~\eqref{eq:ps_link_cons} by assuming the network to have infinite capacity \emph{i.e.} that the operator has to provide the capacity to meet the needs of the users. We then use $\chi(\langle u,v \rangle)$ as a dimensioning metric to estimate required network capacity, later, in the evaluation, we will compare this need under different scenarios.

Our proposed algorithm, given in Algorithm~\ref{alg:swing}, is a variance of the Farthest-First Traversal (FFT), namely \emph{Swing}. Swing objectives is to minimize the distance between a service point and the farthest consumption point. To achieve this objective, the algorithm attempts to spread the service points as wide as possible in the graph, between highly connected nodes and furtherest nodes from the graph center. The difference here is that the node selected next may not necessarily be the farthest node from the current one, but it is the opposite on the scale of connectivity in the graph.
The algorithm takes as input: the network graph, $K$ the number of selections and the $V \times V$ Distance matrix, $\Delta$. Here, \emph{Distance} refers to the minimum cost path between a source and a sink nodes; where \emph{cost} may be represented by various metrics, such as the \emph{hop count}.

First, Swing builds up two sets of candidates for selection: the primary set $\Phi$ and the backup, secondary, set $\Phi'$. Each element in the sets is a vertex-distance pair, $\langle \kappa, \delta \rangle$, the distance part of each pair allows for evaluating the ``goodness'' of the vertex part of the pair; whereby, shorter distance translates into higher goodness. To compose the sets, Swing iterates over all vertices of the graph, steps~\ref{swing:start_candidate}-\ref{swing:end_candidate}; for each vertex $u$, calculates the \emph{connectedness} of its neighbourhood set $V_u$. Connectedness here is defined as the sum of all distances from the neighbour $v \in V_u$ to all other vertices. The neighbour with the best connectedness, \emph{i.e.} minimum total sum of paths length, is added to the primary set $\Phi$, whilst all other neighbours in $V_u$ are added to the backup set $\Phi'$.
Once the sets of candidates are complete, Swing will make $K$ selections of service points from the primary set, then from the backup set if needed, \emph{i.e.} when $K \geq |\Phi|$. When selecting from a candidate set, steps~\ref{swing:k_less_start}-\ref{swing:k_less_end} or steps~\ref{swing:k_more_start}-\ref{swing:k_more_end}, Swing either chooses a candidate with the best connectedness $\Phi[\delta_{min}]$ , if $k$ has an odd value in the range; or, it will select a candidate with the worst connectedness $\Phi[\delta_{max}]$, if $k$ has an even value in the range. Once a candidate is selected \emph{i.e.} added to the selected set $\Psi$, it is removed from the candidacy sets and subsequent selection will be made from the remainder of the sets.  
The performance of the solutions provided by swing is highly dependent on the topology of the graph. Here, we highlight this dependency and parametrise common behaviour of the algorithm.

\subsection{Reference Algorithms}
We compare the performance of Swing against two reference greedy algorithms, namely \emph{Largest First} selection (Pop) in Algorithm~\ref{alg:pop} and \emph{Closest First} selection (Cls) in Algorithm~\ref{alg:cls}. Pop calculates for each vertex $v \in V$ the ratio of the population of the node to the total population in the network, $\rho_v$. It then uses $\rho$ in a probability based selection function $SELECT$. This means the $K$ vertices with largest population have the highest likelihood of selection.
On the other hand, Cls calculates the relative \emph{closeness} centrality of each vertex to the total centrality of all the vertices in the graph, $\rho$. It uses $\rho$ as the selection probability in the function $SELECT$. That is, the $K$ most connected nodes have the highest probability of selection.

Next, we evaluate the performance of our proposed architecture and compare it to that of traditional CDN with DNS redirection. We Also compare our algorithm Swing against the reference algorithms and show its advantages and disadvantages in various scenarios. Moreover, we explore the parameters space and use that to determine best range of $K$ for both origin and edge service points.
\begin{algorithm}[!t]
\caption{$Swing(G(V, A), K, \Delta)$}
\label{alg:swing}
\begin{algorithmic}[1]
\State $\Phi \leftarrow \{\emptyset | \ \forall u \in V\}, \Phi' \leftarrow \{\emptyset | \ \forall u \in V\}, 
\Psi \leftarrow \{\emptyset\}$
\Function {vtx}{$\langle \kappa, \delta \rangle$}
	\State \Return $\kappa$
\EndFunction
\Function {dst}{$\langle \kappa, \delta \rangle$}
	\State \Return $\delta$
\EndFunction
\ForAll {$u \in V$} \label{swing:start_candidate}
%    \State {$\rho \leftarrow \langle \kappa \leftarrow \null, \delta \leftarrow \infty \rangle$}
	\ForAll {$v \in V_u$}
		\State {$\delta_v \leftarrow \sum_{j \in V' \leftarrow V \setminus u}\Delta[v, j]$}
	\EndFor
	    \State {$\Phi \leftarrow \Phi \cup \langle v, \delta_v \rangle, \ \text{where } \delta_v = \min(\delta) , \forall v \in V$}
	    \State {$\Phi' \leftarrow \Phi' \cup \{\ \langle \nu, \delta_{\nu} \rangle \mid \nu \in V_u \setminus v \}$} \Comment {remainder set}
%    \State {$\Phi \leftarrow \Phi \cup \rho$}
\EndFor \label{swing:end_candidate}
	\State {$\Phi \leftarrow \exists_{=1}(\Phi)$} \Comment {ensure set uniqueness}
\If {$K < | \Phi |$} \label{swing:k_less_start}
  \ForAll {$k \in \langle 1 \dots K \rangle$}
      \State {$\epsilon \leftarrow \begin{cases}
      			(k \mod{2}) \ \ \ \Phi[\delta_{min}] \\
                \neg(k \mod{2}) \ \ \ \Phi[\delta_{max}]
      \end{cases}$}
      \State {$\Psi \leftarrow \Psi \cup $\Call{vtx}{$\epsilon$}}
      \State {$\Phi \leftarrow \Phi \setminus \epsilon$}
  \EndFor \label{swing:k_less_end}
\Else \label{swing:k_more_start}
	\State {$\Psi \leftarrow $ \Call{vtx}{$\Phi$}}
    \State {$K' \leftarrow | \Psi| + 1$}
    \ForAll {$ k \in \langle K' \dots K \rangle $}
      \State {$\epsilon \leftarrow \begin{cases}
      			(k \mod{2}) \ \ \ \Phi'[\delta_{min}] \\
                \neg(k \mod{2}) \ \ \ \Phi'[\delta{max}]
      	\end{cases}$}
        \State {$\Psi \leftarrow \Psi \cup $ \Call{vtx}{$\epsilon$}, $\Phi' \leftarrow \Phi' \setminus \epsilon$}
    \EndFor
\EndIf \label{swing:k_more_end}
\State \Return $\Psi$ \label{swing:return_k}
\end{algorithmic}
\end{algorithm}

\begin{algorithm}[!t]
\caption{$Pop(G(V, A), U, K)$}
\label{alg:pop}
\begin{algorithmic}[1]
\State {$\Psi \leftarrow \{\emptyset\}, \ \delta \leftarrow \{0 \mid \forall v \in V\}, \ U_{all} \leftarrow \sum_{v \in V} U_v$}
\ForAll {$v \in V$}
	\State {$\delta_v \leftarrow U_v/U_{all}$}
\EndFor
\State {$\Psi \leftarrow $ \Call{Select}{$from = V, size = K, probability = \delta$}}
\State \Return $\Psi$
\end{algorithmic}
\end{algorithm}
\begin{algorithm}[!t]
\caption{$Cls(G(V, A), K)$}
\label{alg:cls}
\begin{algorithmic}[1]
\State {$\Psi \leftarrow \{\emptyset\}, \ \delta \leftarrow \{0 \mid \forall v \in V\}$}
\ForAll {$v \in V$}
	\State {$\delta_v \leftarrow $ \Call{Closeness}{$v, G$}}
\EndFor
\State {$\Psi \leftarrow $ \Call{Select}{$from = V, size = K, probability = \delta$}}
\State \Return $\Psi$
\end{algorithmic}
\end{algorithm}

\section{Evaluation}
\label{sec:eva}
Here, we analyse the performance of our proposed architecture with respect to the dimensioning parameters and quantify the savings in network resources and improvements in edge-to-edge communications. This is reflected in three parameters: \emph{path length}, \emph{network capacity} and \emph{storage capacity}. Furthermore, we analyse the network capacity gain when delivering through multicast as opposite to unicast in fCDN. We also analyse the performance variability of the placement algorithms within the space of parameters.
As described in Section~\ref{sec:model_and_problem}, our dimensioning problem can be reduced into a multi-objective variance of the K-center problem, which is known to be NP hard. We do not optimise primarily for $K$, but rather explore the parameter space to identify the number of origins and service points that would minimize the joint storage and network requirements. Moreover, we explore the space of catchment intervals for different video lengths and quantify benefits of different catchment intervals for different video lengths.

\subsection{Evaluation Model}
We model our proposed architecture analytically over a realistic network graph from the Internet Topology Zoo~\cite{Knight2011}, namely Geant 2012 $G(V = 37, A = 116)$. We use a synthetically generated content catalogue of 1000 items having a Zipf popularity distribution of exponent $0.8$, following the range of accepted fits for content popularity distribution~\cite{Traverso2013, Cha2009}.
Each item is characterized by: 
\begin{enumerate*}
\item probability of occurrence, drawn from the Zipfian popularity distribution,
\item bitrate value, randomly selected from the range $\{20, 40, 60\}$ Mb/s, and
\item a storage volume randomly selected from the range $\{20, 40, 60\}$ MB.
\end{enumerate*}
The ranges of bit-rates are selected within the common range of bit-rate to support different HD/UHD video formats~\cite{google2017}. Moreover, the popularity distribution is invariant across all nodes in the network. 

To model the subscription rate, we first estimate the network population using the global LandScan population database~\cite{Bhaduri2007}, combined with a standard Voronoi tessellation model associating each potential user to the nearest network node. We consider on average $40\%$ of the population of each node are active users; triggering subscriptions for content in the catalogue. The number of subscriptions per item, in each node, is drawn from the probability of occurrence of the item.

The publication model considers $K = K_o + K_e$ incremental number of publishers in the network, split between origin $K_o = \{2,4,6,8\}$ and edge $K_e = \{2,4,6,8\}$ service points. Origin and edge points are placed at different nodes in the network; where the nodes are selected by one of the selection algorithms, \emph{i.e.} Swing, PoP or Cls. Origins do not have constraint on their storage capacity, hence they are able to cache all the items. Edge points are constrained in storage capacity, where they only cache a subset of the items, while retrieve the rest from origins. To identify which items will be cached in each edge point, we first reverse engineer equation~\eqref{eq:storage_p_def} by estimating first the total storage capacity of $p$, then use a probability-based pseudo random caching policy; whereby, items of higher popularity have a higher likelihood to be selected for caching. The storage capacity can be constrained following different considerations. Here, we assume the capacity placed in each edge point to be relative to the capacity that may be placed in the node with largest population, should it be selected by the optimization algorithm. We base this assumption on the fact that the node with largest population is the node that would require the largest set of items to be cached; thus smaller nodes would require storage resources relative to that of the largest node.
To model traditional CDNs we further introduce an incremental range of LDNS points, $K_d = \{2,4,6,8\}$; the placement of which, is provided by the selection algorithms described in Section~\ref{sec:algo}. Notably, the set of nodes selected to provide LDNS may overlap with the set of publishers. We discuss through the results the importance of this overlap for current CDNs and some of the inefficiencies that may occur if it is not maintained.  

Within the model above, we run $50$ tests of randomized publications and subscriptions for: CDN with DNS redirection, fCDN when delivery is unicast, fCDN when delivery is multicast. Routing is facilitated through Dijkstra's shortest path algorithm, where path length is measured in hop-count. Notably, given the deterministic nature of Swing, only subscriptions are random in each tests. fCDN is modelled for a set of video length values $T = \{900, 1800, 2700, 3600\}$ seconds, and a set of catchment intervals $\tau = \{0.1, 1, 10\}$ seconds.
Next, we present our results observed when modifying different parameters.

\subsection{Results}
\subsubsection{Path Length}
\begin{figure*}[t]
	\centering
	\subfloat [CDN with DNS redirection]{ %
		\includegraphics[width=0.45\linewidth]{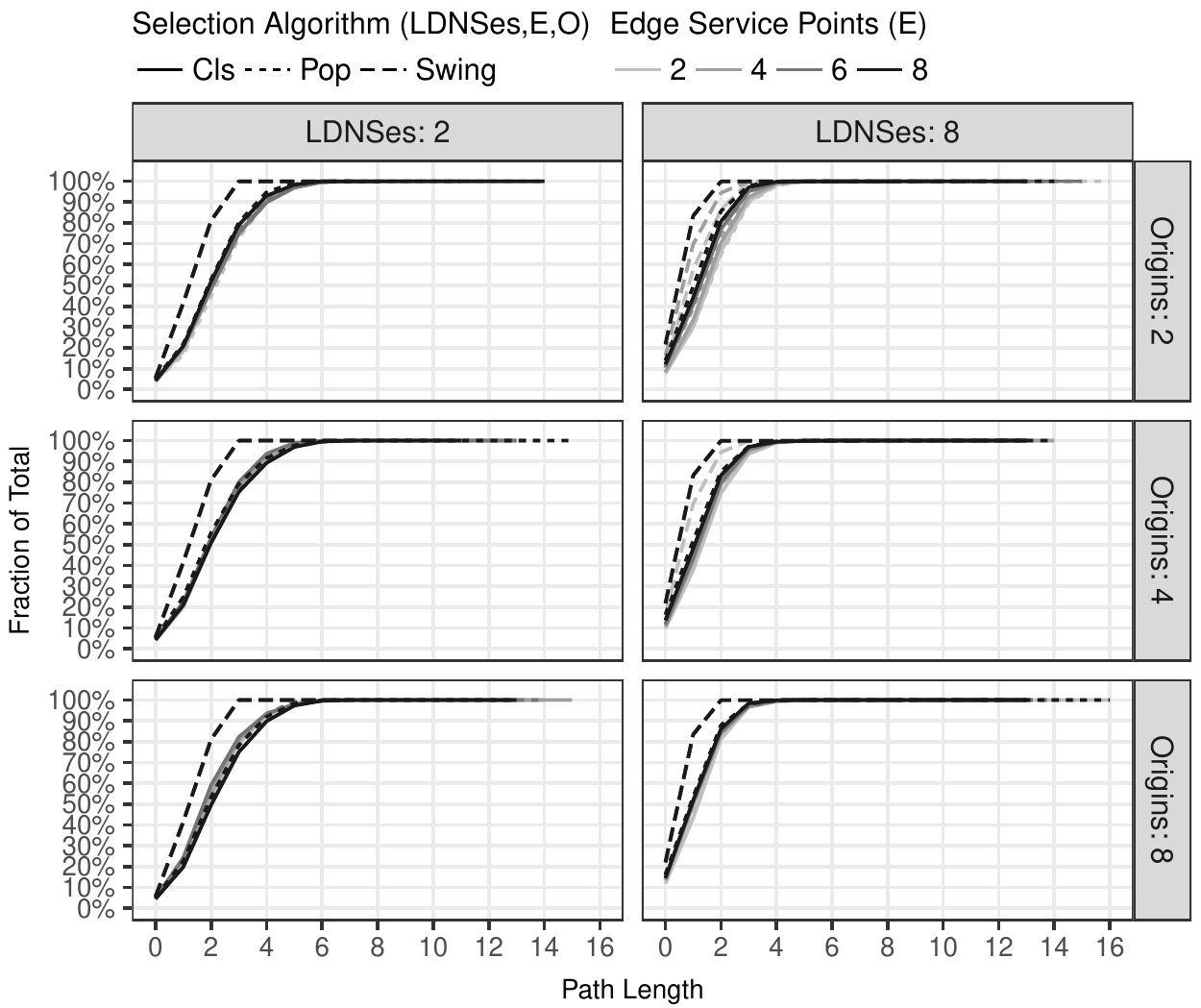}
	}
		\label{fig:ip-pl} \qquad
	\subfloat[flexible CDN (fCDN)]	{%
		\includegraphics[width=0.45\linewidth]{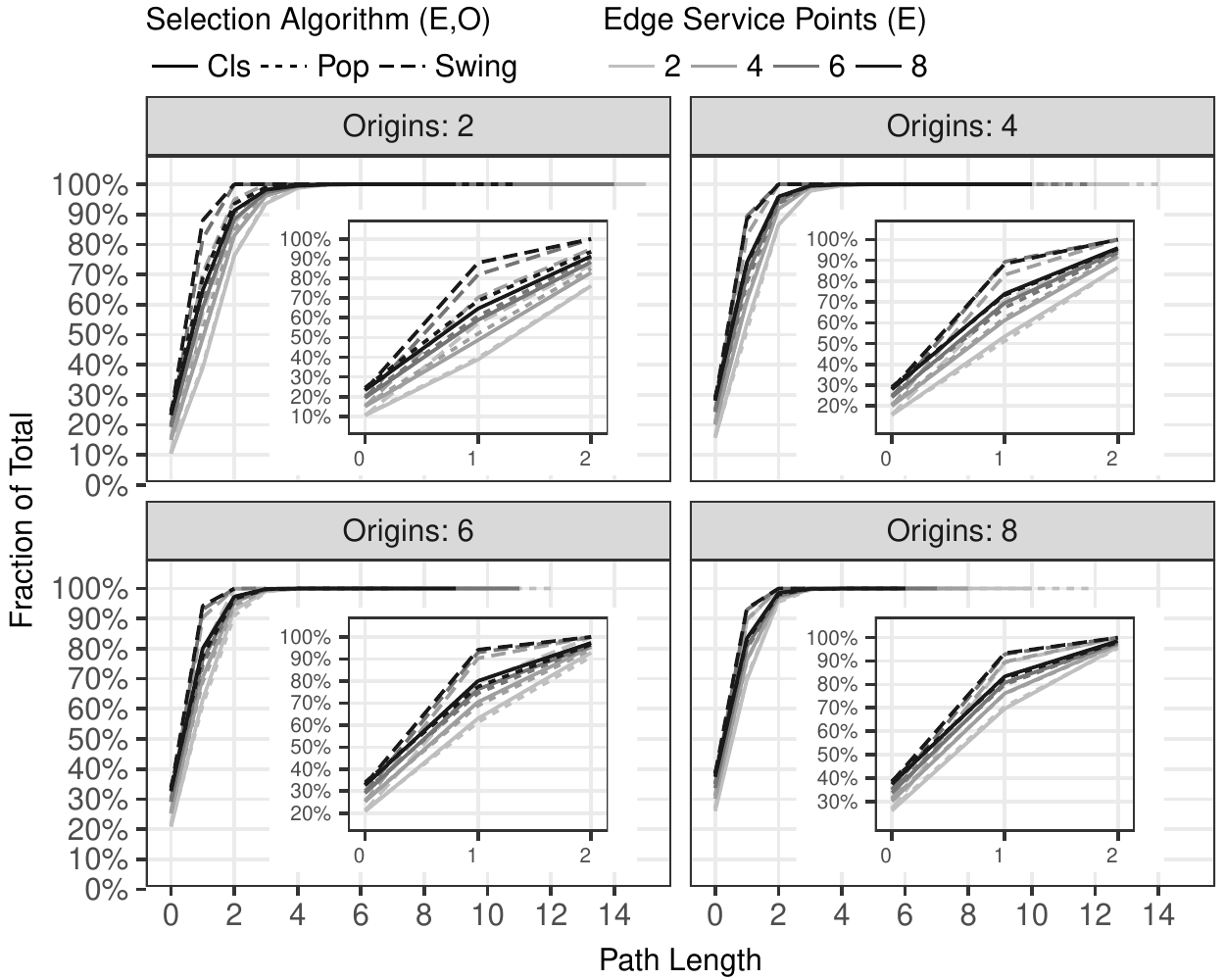}
	}
		\label{fig:icn-uc-pl}
                \\\vspace{0.5em}
	\caption{A comparison between CDN with DNS redirection and fCDN in terms of edge-to-edge path length showing the ECDF of established paths in the Geant graph for increasing number of: origins, edge points and LDNS points, and for all the selection algorithms.}
	\label{fig:pl}
\end{figure*}

Here, we analyse the path length from three perspectives: the converged range of path lengths in CDN and how it compares to fCDN, variations in this convergence for CDN in response to varying the number and position of LDNS points, and the performance of the placement algorithms. Figures~\ref{fig:ip-pl} and~\ref{fig:icn-uc-pl} shows the empirical cumulative distribution function of all the paths in the CDN with DNS reduction and in fCDN respectively. Due to complexity constraints we show IP path length for the minimum and maximum number of LDNSes, $\{2,8\}$. The results show that when the number of LDNS points is small, $2$, approximately $5\%$ of the paths has a zero-length. Those reflect localized demands, where the node has a service point (origin or edge). The highest path length is $15$ hops and approximately $90\% - 92\%$ of the paths are $4$ hops or less. Notably, increasing the number of  service points alone, \emph{i.e.} without a comparable increase in the number of LDNS points, does not introduce a noticeable decrease in the path length. 
It is only when the number of LDNS points increases, $8$, that a reduction is witnessed. In this case, between $10\%$ and $20\%$ of paths are of zero length, whereas the maximum path length varies between $14$ and $16$ hops depending on $K_o$ number of origins. Convergence varies as well; when $K_o$ number of origins is $2$ the variation between the different algorithms is distinguishable. Nonetheless, proximately $90\%$ of paths are of $3$ hops. When $K_o = \{6,8\}$, around $87\%$-$88\%$ of the paths are of $2$ hops.
This behaviour is purely due to the inefficiency in mapping requests based on DNS redirection, where the service point is selected closest to the LDNS but not necessarily closer to the consumption point. This means a node might have a service point placed at its site and has a cached copy of the item of interest, but if the LDNS is remotely located and adjacent to a farther service point, requests of the node will be mapped to the remote service point rather than itself. This leads to a rigid constraint on co-locating LDNS points with edge or origin service  points when dimensioning a CDN network, , to avoid inefficient request mapping.

In contrast, we observe in Figure~\ref{fig:icn-uc-pl} that path lengths in our proposed fCDN are considerably shorter that their counterpart in traditional CDN. For a small number of origins $K_o = 2$, $20\%$ to $30\%$ of paths are of length zero;  for larger $K_o = 8$, $30\%$ to $40\%$ of paths are of length zero. This indicates around $2-4$ fold increase in the number of localized demands due to accurate request mapping. Maximum path lengths also show a considerable reduction from $14$ hops in best case to $10 - 12$, when $K_o = \{6,8\}$. Moreover, when the number of service points is small, $K_o = 2$ and $K_e = 2$, approximately $90\%$ of paths are of $2$-$3$ hops. As the values of $K_o$ and $K_e$ increases to $8$ each, more than $95\%$ of paths reduces to $2$ hops.
Notably, when comparing the algorithms behaviour, we observe that Swing facilitates significantly shorter paths compared to Pop or Cls. On the left hand side of Figure~\ref{fig:ip-pl}, When Swing is used to place services, approximately $99\%$ of paths are of $3$ hops with the longest path being of $7$ hops as apposite to $13$ and $14$ hops by Pop and Cls respectively. On the right hand side of the same figure, approximately $99\%$ of paths associated with Swing are of $2$ hops with maximum path length between $5$ and $15$, depending on $K_e$ number of surrogates. Similarly, in Figure~\ref{fig:icn-uc-pl}, over $99\%$ of paths associated with Swing are of $2$ hops; and the longest path is of $4$ hops. This shows the superiority of Swing over the reference algorithms in consistently reducing the path length for both CDN and fCDN.

%-----------------------------
%-----------------------------

\subsubsection{Network Capacity}
\begin{figure*}[tb]
	\centering
	\subfloat[CDN with DNS redirection]
	{
		\includegraphics[width=0.45\linewidth]{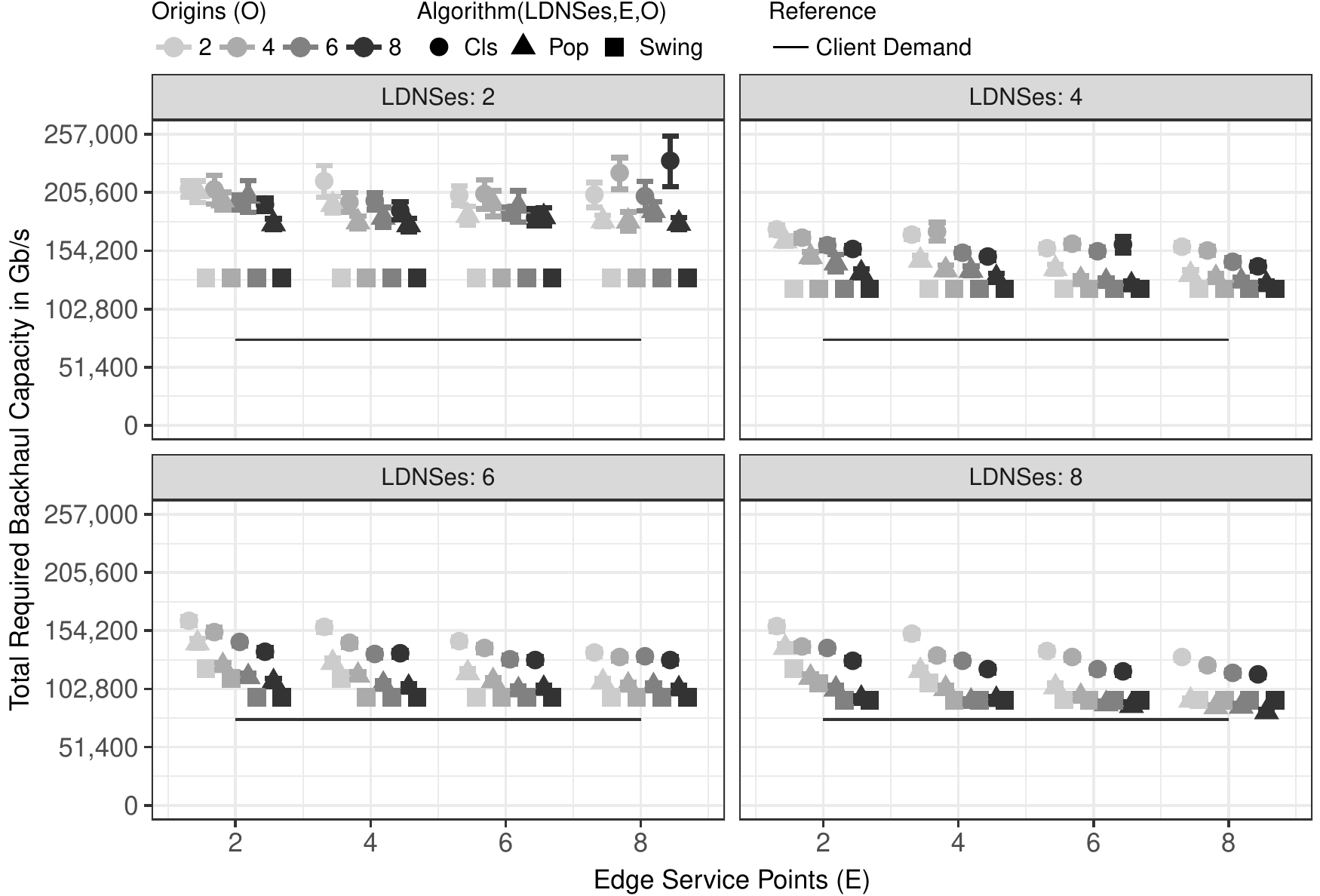}
	}
		\label{fig:ip-capacity}\qquad
	\subfloat[flexible CDN (fCDN) unicast and multicast mode]
	{
		\includegraphics[width=0.45\linewidth]{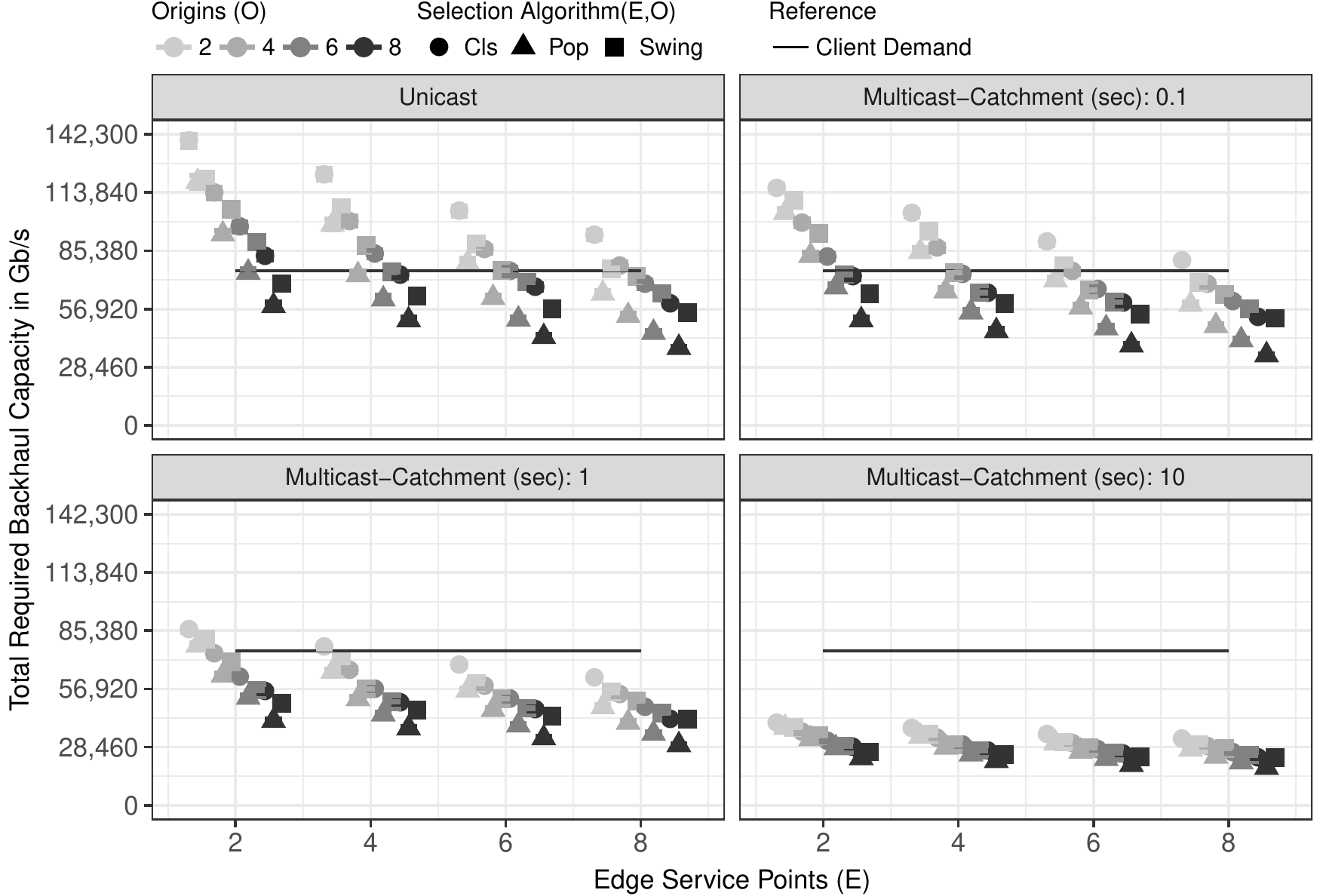}
	}
		\label{fig:icn-uc-capacity}
         \\\vspace{0.5em}
	\caption{A comparison between CDN with DNS redirection, unicast fCDN and multicast fCDN, in terms of backhaul required capacity showing the total required capacity to accommodated the offered demand of 70 Gb/s in the Geant graph for increasing number of: origins, edge points and LDNS points, and for all the selection algorithms.}
	\label{fig:net-capacity}
\end{figure*}

Here we evaluate the network capacity required to accommodate the offered demand, for an increase number of origins $K_o = \{2,4,6,8\}$ and increasing number of edge service points $K_e = \{2,4,6,8\}$. Moreover, we also tune $K_d = \{2,4,6,8\}$ number of LDNSes for traditional CDN. Through this tuning of parameters, we argue a mixture of $Ks$ that results in minimum backhaul capacity requirements. Figures~\ref{fig:ip-capacity} and~\ref{fig:icn-uc-capacity} shows the required backhaul capacity to accommodate a total theoretical  demand of $70$ Gb/s, in both CDN and fCDN respectively. For CDN with DNS redirection, the backhaul capacity is directly tied to the number of LDNSes placed in the network. For a small number of DNSes, $K_d = 2$, tuning the other $K_o$ and $K_e$ has negligible impact on the required backhaul capacity; however, as $K_d$ increases, the impact of increasing $K_e$ and $K_o$ increases. In all cases the average required backhaul is approximately between $100$ and $200$ Gb/s, making  $1-2$ folds of the demand.

In contrast, the backhaul capacity of fCDN is highly dependent on the number of service points $K_e$ and $K_o$. Figure~\ref{fig:icn-uc-capacity} shows the backhaul capacity required in fCDN for the two dissemination modes: unicast and multicast. For the latter, we assume a fixed video length of $900$ seconds and we vary the catchment interval in the period $\{0.1, 1,10\}$. For a small number of edge points, $K_e =2$, the backhaul capacity required for unicast is approximately $115$ Gb/s, averaged for all algorithms, \emph{i.e.} $2$ folds of the demand. However, as the number of origins increases to $K_o =8$, the backhaul capacity drops to $\approx 65 $ Gb/s that is lower than the offered demand of $70$ Gb/s. Similar reduction is observed when fixing the number of origins to $K_o = 2$ and increasing $K_e$ gradually from $2$ to $8$ where the backhaul traffic drops from $\approx 115$ Gb/s to $\approx 70$ Gb/s. This reduction in backhaul traffic is a result of localising substantial portion of the offered demand, due to accurate request mapping to the nearest service point. Backhaul traffic can further be reduced when delivery is multicast, depending on the catchment interval $\tau$ that can be achieved. When $\tau = 0.1$ second, an approximately $15\%$ reduction in backhaul traffic is observed; and, when $\tau = 1$ second, the reduction ratio increases to $\approx 50\%$. Notably, as the catchment interval increases, the difference in performance between the algorithms decreases. Notice that a catchment interval of $10$ seconds is unrealistic as it introduces substantial delay to the down time and affects the users' perceived  Quality of Experience (QoE). Nonetheless, we include this value to give an indication for the further improvements that can be achieved with realistic $1 \leq \tau \leq 10$ values.

Notably, the results also show that the difference in capacity requirements is small when $K_e =2$ and $K_o = 8$ as opposite to the combination of $K_e = 8$ and $K_o = 2$. This indicates that given a number of publishers  $K =  K_e + K_o$, increasing the caching points as opposite to increasing the advertising points, with the content not cached, does not largely affect the backhaul capacity. This translates into possible savings in the storage capacity, without added cost on the network, as we will see later in Section~\ref{sec:eva-sc}.
%-----------------------------
%-----------------------------
\subsubsection{Multicast Gain}
Here, we analyse the network capacity gain when delivering content over multicast trees. We show how the multicast gain is affected by the catchment interval and the video length, as described in~\eqref{eq:mc_T_over_t}. 
 The multicast gain has been evaluated for catchment intervals $\tau \in \{0.1, 1, 10\}$ for an average video length $T=900$ seconds. For a catchment interval of $0.1$second, the multicast gain is $\approx 1.25$; whereas, for $\tau = 1$ second the gain is between $1.4$ and $1.8$. The results here are aligned with the approximately $2$ fold reduction in backhaul capacity, when the catchment interval is $1$ second. Higher gains can be realized with higher $\tau$
When considering different video lengths $T \in \{1800,2700,3600\}$ seconds. The results indicates that the average backhaul capacity and gain maintains similar profile to that of $900$ seconds. However, for a fixed $\tau$, videos of higher length will offer smaller capacity gain. This is expected as the average number of subscriptions decreases for longer videos, given the homogeneous Poisson distribution, described in Section~\ref{sec:model_and_problem}.
%%%-----------------------------
%\begin{figure*}[t]
%	\centering
%	\subfloat[Multicast gain]{
%			\includegraphics[width=0.45\linewidth]{icn_multicast_gain.pdf}
%	}
%	\label{fig:icn-mc-ga}
%	\subfloat[Required Backhaul capacity using multicast]{
%			\includegraphics[width=0.45\linewidth]{icn_multicast_variation_capacity_gain.pdf}
%	}
%	\label{fig:icn-mc-ga-var}
%	\caption{(a), (b) Multicast traffic analysis in Geant 2012 network\mjreed{I do not understand b, what is the graph on the right compared to the results in (a), maybe cut this. Could even cut this complete figure and just state average multicast gains to save space.}}
%	\label{fig:net-gain}
%\end{figure*}
%%%-----------------------------
\subsubsection{Storage Capacity}
\label{sec:eva-sc}
\begin{figure}[tb]
\centering
\includegraphics[width=1\columnwidth]{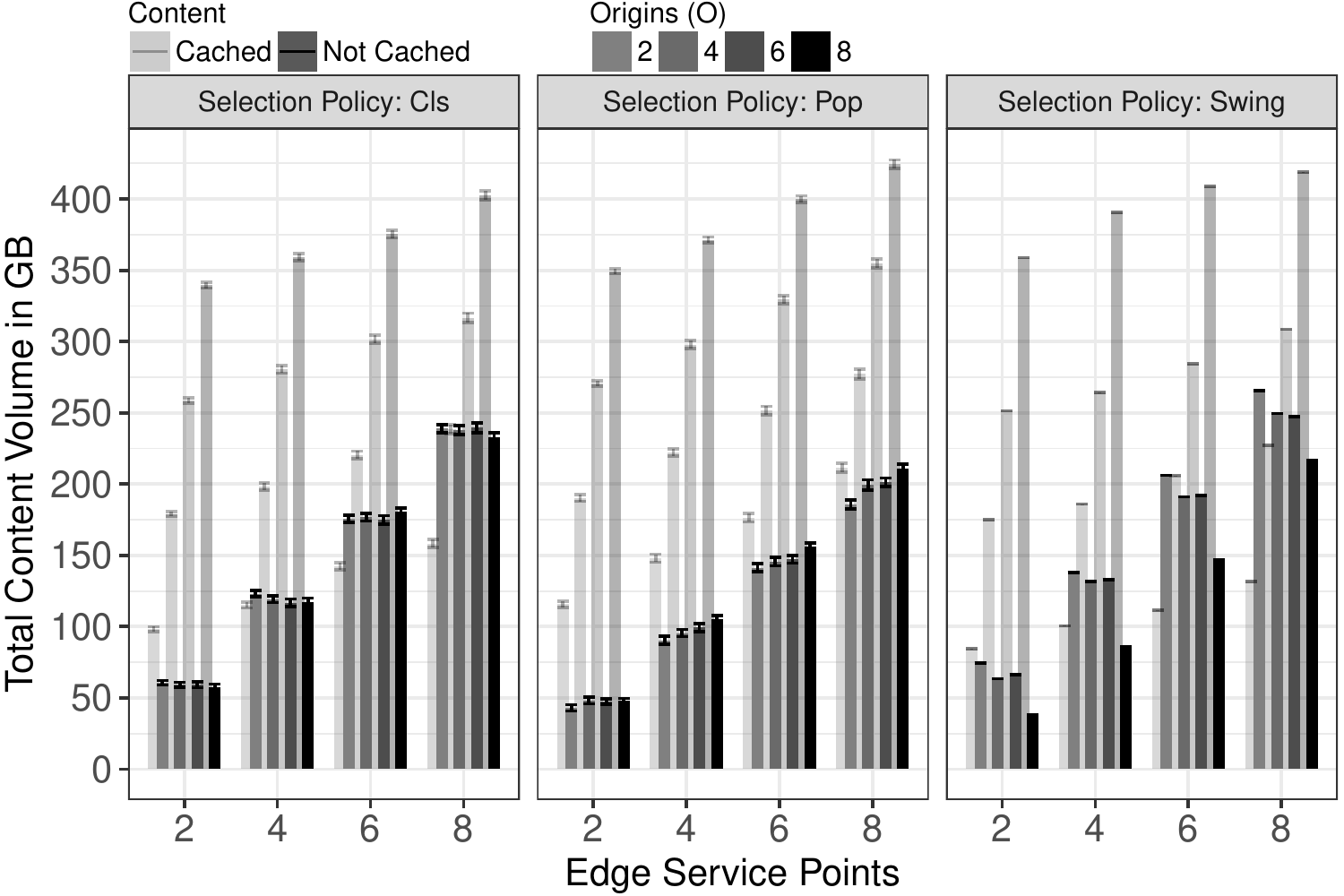}
\caption{A comparison between the selection algorithms Swing, Pop and Cls in terms of storage capacity in the Geant graph when increasing the number of origins for a fixed number of edge points, as opposite to increasing edge points for a fixed number of origins}
\label{fig:sc}
\end{figure}

Here, we analyse the storage capacity required in the network when increasing the number of edge points  $K_e$ for a fixed number of origins $K_o$; and, when increasing $K_o$ for a fixed $K_e$. We illustrate this by calculating the theoretical total volume of all items; that is, the  total volume if all items are stored in all selected nodes, \emph{i.e.} $K = K_o$. We then compare the ratio of volume of all items that have been cached and advertized with the ratio of volume that has been advertized, but not been cached. 
%We show through this comparison the different storage requirements and how they relate to backhaul capacity requirements.
Figure~\ref{fig:sc} illustrates the storage requirements for $K_e \in \{2,4,6,8\}$ and $K_o \in \{2,4,6,8\}$. For a small number of edge points $K_e=2$ and large number of origins $K_o=8$, only $25\%$ of the total volume is cached. In contras, when  $K_e= 8, K_o = 2$, approximately $60\%$ of the volume is advertized but not cached, while only $\approx 40\%$ of the volume is cached. When comparing this result with that of Figure~\ref{fig:icn-uc-capacity} for backhaul capacity, we find that the difference in cached vs. not cached does not have a significant impact on the backhaul traffic, as long as the content is advertized by all nodes. This means, advertising the availability of content without necessarily caching it is sufficient enough to localise the offered demand. This is because edge points retrieve content from the origins always through orchestrated multicast, which reduces the amount of backhaul traffic to substantially small values. 
%-----------------------------
%-----------------------------

From the results presented so far, we can draw two general views: first, increasing content visibility through larger number of edge points that advertise the content but not cache it, while dimensioning a low number of origins, allows for reducing both storage and backhaul capacity requirements, while accommodating the offered load. Second, backhaul capacity can further be reduced and capacity gain can be obtained, when delivering over multicast trees rather than unicast paths. The gain in backhaul capacity due to multicast is subject to the variation of the catchment interval, $\tau$. A significant $50\%$ reduction can be achieved, when setting $\tau = 1$ second. 

\section{Conclusion}
\label{sec:conclusion}
Flexible CDN architectures are critical in meeting the expected growth of edge communications in 5G+ networks. Here, we proposed a novel and flexible CDN (fCDN) architecture that eliminates two of the major inefficiencies in existing CDNs, namely the suboptimal request mapping due to DNS redirection and the suboptimal routing by the transport overlay. Instead, our proposed fCDN facilitates anycast request mapping and multicast routing directly in the network, through a Publish/Subscribe matching function embraced from emerging research in ICN. Furthermore, to facilitate service placement, we formulated the problem of service placement as a  variance of the K-center problem; and, proposed a novel greedy algorithm,named Swing, to solve it. We evaluated the performance of our architecture, and the proposed algorithm, with respect to request mapping using synthetically generated demands over realistic network topology. Our evaluation results showed that increasing the number of advertising edge points, for a fixed number of origins is preferable over increasing the number of origins for a fixed number of edges.

\section*{Acknowledgment}
This work was supported by the European Union funded H2020 ICT project POINT, under contract
No 643990

%\ifCLASSOPTIONcaptionsoff
%  \newpage
%\fi



\bibliographystyle{IEEEtran}
%\bibliography{ref}


%\end{IEEEbiography}


%\newpage



%\vfill


\end{document}